\documentclass[a4paper,12pt]{article}

\usepackage{arxiv}

\usepackage{scalefnt}

\usepackage[utf8]{inputenc} 
\usepackage[T1]{fontenc}    
\usepackage{url}            
\usepackage{booktabs}       
\usepackage{amsfonts}       
\usepackage{nicefrac}       
\usepackage{microtype}      
\usepackage{doi}

\usepackage{amsmath}
\usepackage{tabularx}
\usepackage{graphbox,graphicx}
\usepackage[export]{adjustbox}
\usepackage[margin=10pt,font=small,labelfont=bf]{caption}
\usepackage{subcaption}
\usepackage{siunitx}
\usepackage{array, threeparttable, booktabs,  caption}
\usepackage{dcolumn}
\newcolumntype{d}[1]{D{.}{.}{#1}}

\usepackage{longtable}
\usepackage{threeparttable}
\usepackage{threeparttablex}
\usepackage{tikz}
\usetikzlibrary{shapes,decorations,arrows,calc,arrows.meta,fit,positioning}
\tikzset{
    -Latex,auto,node distance =1 cm and 1 cm,semithick,
    state/.style ={ellipse, draw, minimum width = 0.7 cm},
    point/.style = {circle, draw, inner sep=0.04cm,fill,node contents={}},
    bidirected/.style={Latex-Latex,dashed},
    el/.style = {inner sep=2pt, align=left, sloped}
}

\usepackage{float}

\usepackage[bottom]{footmisc}
\usepackage{fancyhdr} 	
\usepackage{enumitem}
\usepackage{chngcntr}	

\DeclareFontFamily{U}{mathx}{\hyphenchar\font45}
\DeclareFontShape{U}{mathx}{m}{n}{<-> mathx10}{}
\DeclareSymbolFont{mathx}{U}{mathx}{m}{n}
\DeclareMathAccent{\widebar}{0}{mathx}{"73}

\newcommand{\RNum}[1]{\uppercase\expandafter{\romannumeral #1\relax}}

\title{The Importance of Quality in Austere Times: University Competitiveness and Grant Income}

\author{ \href{https://orcid.org/0000-0001-6638-9707}{\includegraphics[scale=0.06]{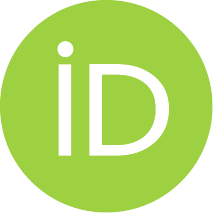}\hspace{2mm} Ye Sun}\thanks{We extend our gratitude to Hanna Hottenrott and Chirantan Chatterjee for their feedback on the paper. We would also like to thank participants for comments at the SPRU Freeman Seminar, the 2022 PET conference and the 9th ZEW/MaCCI Conference on the Economics of Innovation and Patenting. Vito Latora acknowledges support for this research by the Leverhulme Trust Research Fellowship “CREATE: the network components of creativity and success”.  } \\
    {\small 
	Department of Computer Science }\\ 
     {\small University College London} \\ 
     {\small United Kingdom } \\
	 {\small \texttt{ye.sun@ucl.ac.uk} } \\
	\And
	\href{https://orcid.org/0000-0002-9858-4889}{\includegraphics[scale=0.06]{orcid.pdf}\hspace{2mm} Athen Ma} \\
    {\small 
    School of Electronic Engineering and Computer Science } \\
	 {\small Queen Mary University of London} \\
	 {\small London E1 4NS, United Kingdom} \\
	 {\small \texttt{a.ma@qmul.ac.uk} } \\
	\And
    \href{https://orcid.org/0000-0003-3935-1679}{\includegraphics[scale=0.06]{orcid.pdf}\hspace{2mm} Georg von Graevenitz} \\
    {\small 
    School of Business and Management} \\
	 {\small Queen Mary University of London} \\
	 {\small London E1 4NS, United Kingdom} \\
	 {\small \texttt{g.v.graevenitz@qmul.ac.uk}} \\
	\And
    \href{https://orcid.org/0000-0002-0984-8038}{\includegraphics[scale=0.06]{orcid.pdf}\hspace{2mm} Vito Latora} \\
    {\small
    School of Mathematical Sciences} \\
	 {\small Queen Mary University of London} \\
	 {\small London E1 4NS, United Kingdom} \\
	 {\small \texttt{v.latora@qmul.ac.uk} }\\
}

\date{This manuscript was compiled on \today}

\begin{document}

\maketitle


{\normalsize
\begin{abstract}
	After 2009 many governments implemented austerity measures, often restricting science funding. Did such restrictions further skew grant income towards elite scientists and universities? And did increased competition for funding undermine participation? UK science funding agencies significantly reduced numbers of grants and total grant funding in response to austerity, but surprisingly restrictions of science funding were relaxed after the 2015 general election. Exploiting this natural experiment, we show that conventional measures of university competitiveness are poor proxies for competitiveness. An alternative measure of university competitiveness, drawn from complexity science, captures the highly dynamical way in which universities engage in scientific subjects. Building on a data set of 43,430 UK funded grants between 2006 and 2020, we analyse rankings of UK universities and investigate the effect of  research competitiveness on grant income.  When austerity was relaxed in 2015 the elasticity of grant income w.r.t. research competitiveness fell, reflecting increased effort by researchers at less competitive universities. These scientists increased number and size of grant applications, increasing grant income. The study reveals how funding agencies, facing heterogeneous competitiveness in the population of scientists, affect research effort across the distribution of competitiveness.
\end{abstract}
}

\keywords{Science of science $|$ Research competitiveness $|$ Grant funding $|$ Austerity \\
$\qquad$ \\}

\newpage

\linespread{1.1}

\setcounter{page}{1}
\pagestyle{plain}

\section{Introduction}

There appear to be strong secular trends towards greater asymmetry in science: increased grant sizes \cite{BS15}, clubbing together of leading universities on grant applications \cite{MML15}, increasing citation inequality \cite{NA21} leading to disputes about funding allocation \cite{wpW19}. Furthermore, doing good science and profiting from it is getting harder \cite{bJ09,ABD20}. Adding to this gloomy picture, the global financial crisis (GFC) of 2008 led to real terms reductions or freezes in funding for scientific research in many countries \cite{LGS18,JSS19}. Success rates at many science funding bodies also declined significantly after the GFC \cite{JSS19}. Did the GFC act to further deepen the divide between the haves and the have-nots in science? Status effects such as the Matthew effect \cite{rkM68,ASW14,BVR18}, but also contest theory \cite{MS06,FNS20,OS20} suggest austerity should have resulted in a concentration of grant incomes. After 2015 austerity in science funding was unexpectedly relaxed in the UK \cite{GP16}. Contest theory suggests less competitive universities would respond by applying for more funding, but recent work on the Matthew effect \cite{BVR18} suggests they might not. In the UK grant applications submitted by less competitive universities grew and their grant incomes  increased.

An analysis of how austerity affected universities' grant incomes must account for heterogeneity in university research competitiveness. Hitherto publication or grant income data are used to measure competitiveness~\cite{JJ18,VBM18,GWB19}. This ignores that publication data can only reveal excellence with a time lag to allow accumulation of citations. Use of Journal Impact Factors is now deprecated \cite{dora, LM15, dec}. Where science is sufficiently dynamic \cite{MB04,KPH14} publication based metrics will reflect past competitiveness, but mismeasure current competitiveness. This problem could be exacerbated by funding cuts. Measures of grant income disbursed after peer review are forward looking indicators. Unfortunately, measures of aggregate grant income may also be confounded, if science is dynamic.  Aggregating income gives equal weight to all subjects, growing or declining, complex or not. Simple aggregates might rank two universities equally, even if one is focused only on growing, new subjects while the other is focused only on declining, older subjects. Therefore, a  metric based on grant income data that captures growth and decline of science subjects is highly needed. 

Our contribution is threefold: i) drawing on recent methods developed in complexity science~\cite{HH09,cH21,tacchella2012new}, we propose a metric of research competitiveness that ranks universities based on the variety and complexity of their research subjects; ii) we show that both university research competitiveness and subject complexity 
are highly dynamic over time; iii) we find that austerity concentrated grant income among the UK's more research competitive universities and that once austerity was relaxed, this was reversed. A comparison of our proposed metric to conventional aggregates of grant income  demonstrates that these do not reflect the effects of austerity at all. These findings are derived using grant income data from UK research councils at the subject-grant level for the period 2006-2020. 

\section*{Results}

\begin{figure*}[ht!]
    \includegraphics[width=16.5cm]{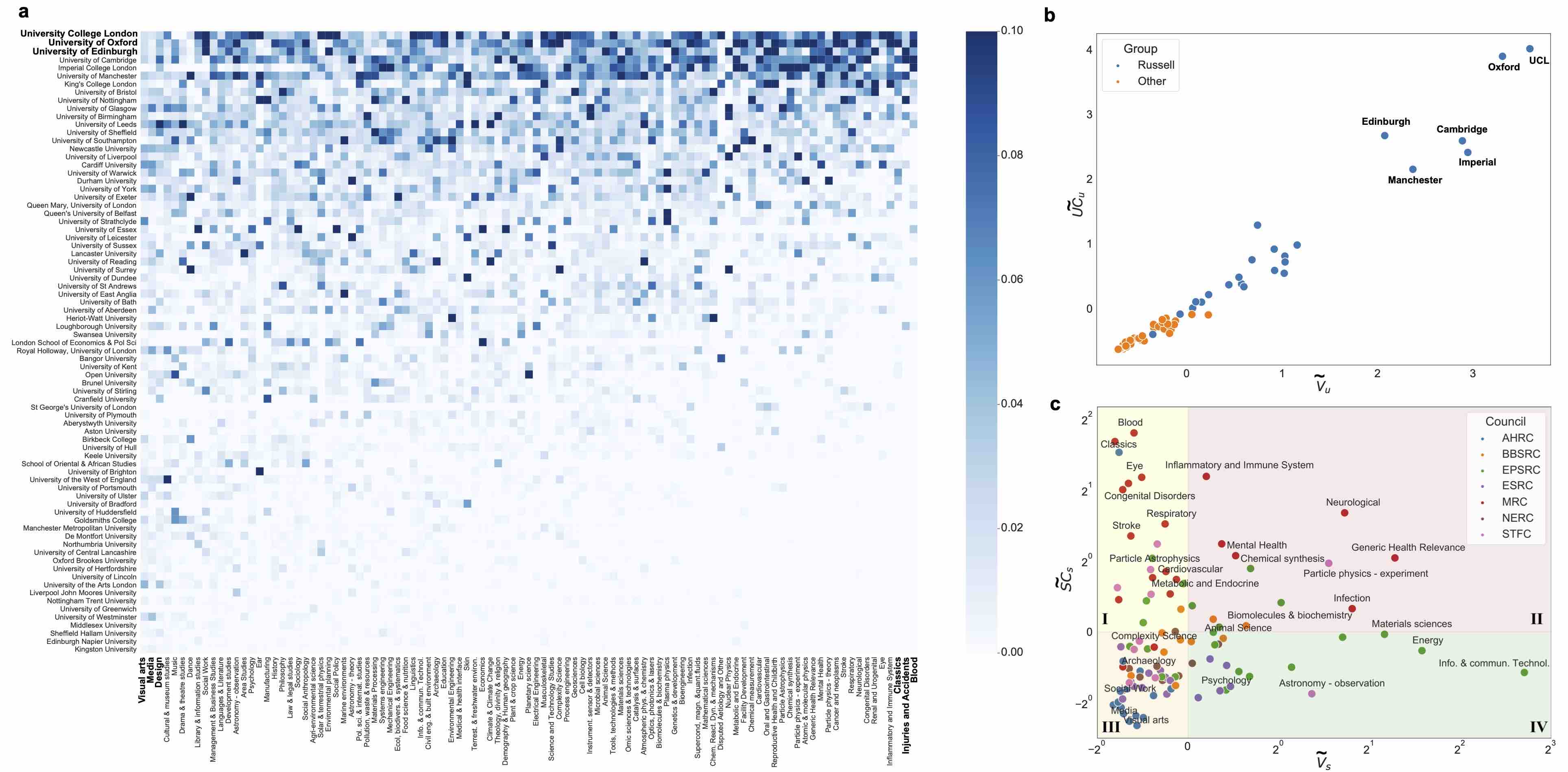}
    \captionsetup{justification=justified}
    \caption{\footnotesize{\textbf{Ranking universities and subjects} \textbf{(a)} The weighted university-subject bipartite $M_{u,s}$ matrix for 15-y period from $2006$ to $2020$. Moving away from the origin university competitiveness (UC$_u$) and subject complexity (SC$_s$) increase. Cell color reflects the proportion of the research funding that a university received in a given subject. Entries across universities sum to $1$ for each research subject. \textbf{(b)} Correlation between UC$_u$ and total awarded funding V$_{u}$. \textbf{(c)} Complexity-Value diagram divided into $4$ regions by average $\left \langle \text{SC}_s \right \rangle$ and funding value $\left \langle \text{V}_{s} \right \rangle$. Nodes of similar color are pulled from the same research council. Note that  $\widetilde{\text{SC}}_s$, $\widetilde{\text{V}}_s $, $\widetilde{\text{UC}}_u $ and $\widetilde{\text{V}}_u $ are z-scores.}}
    \label{fig:overall_feature}
\end{figure*} 

Expanding the range of applications of  economic complexity metrics ~\cite{HH09,tacchella2012new} to university grant funding we construct 
{\em University Competitiveness} (UC) and {\em Subject Complexity} (SC) metrics. These two metrics are based on grant income data, and are derived from a bipartite weighted university-subject network which describes 
the amount of funding received by each university in each research field (\hyperref[sec:m]{Methods}). This approach to measuring complexity exploits the idea that research fields in which only few universities attract funding are likely to be more complex, whilst those with many funded grants are less so. Starting from the assumption that all fields are equally complex and all universities equally competitive the algorithm uses information on grant income by university and field to iteratively rank all universities and fields.

Fig.~\ref{fig:overall_feature}a presents a weighted university-subject bipartite network, where UK universities and subjects have been sorted in descending order, from top to bottom (UC$_u$) and from right to left (SC$_s$), respectively. Here $_u$ stands for university, and $_s$ for subject. The resulting matrix has an approximate triangular shape, indicating that the most research competitive universities obtain significant grant shares in almost all academic disciplines, while the diversification across subjects declines for less competitive universities. Leading universities tend to maximally diversify their research fields in the funding system, rather than to specialise. In particular, they are competitive in complex research subjects, which are on the right in Fig.~\ref{fig:overall_feature}a. According to this aggregate ranking, the top three most competitive universities between 2006 and 2020 were \emph{University College London}, \emph{University of Oxford} and \emph{University of Edinburgh}, respectively, in line with the expectation that these universities have capabilities and resources to conduct research in  any field of science. This matrix presents a similar pattern of specialisation to those documented in the economic complexity literature for countries and exports \cite{HH09,MFT19} or regional economies and employment \cite{FM21}.

The ranking of subjects by complexity indicates the most complex research subjects are found in the medical sciences (such as \emph{Blood}, \emph{Injuries and Accidents}, \emph{Inflammatory and immune system}) and in particle physics (Fig.~\ref{fig:overall_feature}a). These subjects are more likely to have high barriers to entry, often requiring multi-dimensional resources, such as highly advanced equipment and facilities, trained specialists and significant research investment. Subjects from the arts and humanities (e.g., \emph{Visual arts}, \emph{Media} and \emph{Design}), the social sciences (e.g., \emph{Management and business studies}, \emph{Social work}) are identified as the least complex as most research universities have the capacity to obtain grant funding in these subjects.

Total grant funding is often viewed as an indicator of research competitiveness \cite{hfM17,GWB19}. Fig.~\ref{fig:overall_feature}b compares university grant funding with UC$_u$.  The figure reveals that UC$_u$ is strongly correlated with total awarded funding V$_{u}$, and that Russell Group universities obtain more grant income than other UK universities \cite{MML15}. Note, Figure \ref{fig:rank} in the \hyperref[sec:suppinf]{SI Appendix} shows that rankings based on grant income differ from rankings based on UC$_{u}$. Next, Fig.~\ref{fig:overall_feature}c compares subject funding levels with SC$_s$. Comparing subject level grant income with the SC$_s$ we divide the space into $4$ regions by average subject complexity, $\left \langle \text{SC}_s \right \rangle$, and funding value, $\left \langle \text{V}_{s} \right \rangle$. We normalise SC$_s$ and V$_s$ by z-score transformation \cite{cH21}. These indices are displayed on a binary logarithmic scale. 

Research subjects in Quadrant \RNum{2} (right upper corner) are concentrated, high-investment subjects such as \emph{Neurology}, \emph{Particle physics} and \emph{Cancer}. Conversely, the subjects in Quadrant \RNum{3} at the left bottom require low-investment. These include \emph{Visual arts} and \emph{Media}. Subjects from Quadrant \RNum{1} are exclusively researched at leading universities but require less funding; subjects in Quadrant \RNum{4} are well-funded and researched in many universities, indicating lower complexity. The figure reveals that medical research currently tends to be more complex than the engineering and physical sciences and that both are more complex than social sciences and the humanities.

\subsubsection*{Dynamics of Scientific Fields and Universities}

\begin{figure*}[ht!]
    \includegraphics[width=16cm]{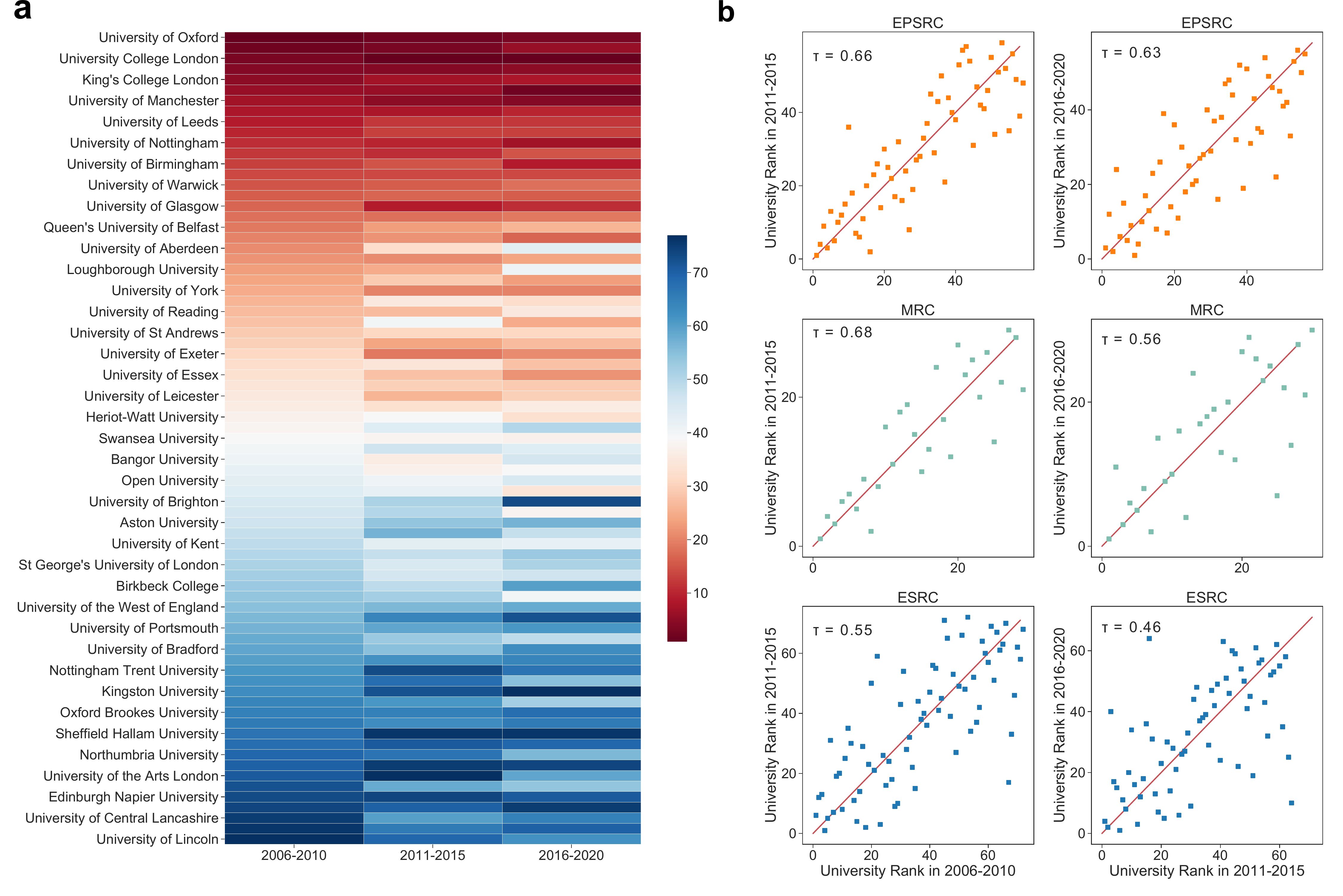}
    \captionsetup{justification=justified}
    \caption{\textbf{Dynamics of UK universities' research competitiveness.} \textbf{(a)} Dynamics of UC$_{p}$ rankings for selected universities for three 5-y \emph{p}eriods, corresponding to successive UK governments. Full rankings are provided in the \hyperref[sec:suppinf]{SI Appendix}. Universities ranked in descending order by UC$_{p}$ for 2006-2010. \textbf{(b)} Dynamics of UC$_{f,p}$ rankings at research council level: each point indicates the ranking of a university in successive periods. The overall variation in rankings is measured by the reported values of of the Kendall’s $\tau$ coefficient.}
    \label{fig:evol_uni}
\end{figure*}

Does the complexity and significance of subjects, as revealed by funding levels, vary over time? Previous literature documents that science throws up new, fast-growing topics with great regularity \cite{MB04,KPH14}. The availability of data-sets comprised of scientific papers covering many disciplines has led to advances in the delineation of scientific fields ~\cite{DGSA18} and in characterising their life cycle \cite{SBWTS21}. Here we exploit the subject classification provided in our data to analyse the dynamics of funding allocations to universities and subjects. Our results, illustrated in Fig.~\ref{fig:evol_uni}, show that science is also highly dynamic when studied through the lense of grant funding. One implication is that rankings of university research competitiveness based on publications and citations are likely to be unreliable as measures of present or future research competitiveness.

We study dynamics of UC$_{u,p}$ and SC$_{s,p}$ over three 5-y time windows, separated by the UK general elections of 2010 and 2015. Here $_p$ stands for period.  Fig.~\ref{fig:evol_uni}a illustrates the evolution of university rankings based on UC$_{u,p}$. Ranking of the most research competitive universities are comparatively stable, while ranks of other universities can vary significantly. For instance, \emph{University of Essex}, \emph{University of Hull} and \emph{Northumbria University} all improved their positions by more than $10$ ranks (Fig.~\ref{fig:Evolution_Uni_Sub_Emphasis}a). These universities improved because they do research in subjects that have grown increasingly complex, and they have secured a greater share of funding in these subjects. \emph{University of Essex} improved its ranking, by developing the capability to conduct research in more complex fields such as \emph{Blood}, and also the volume of funding it has received in certain areas has increased significantly, such as \emph{Library \& information studies} (from $0\%$ to $58\%$) and \emph{Demography \& Human geography} (from $4\%$ to $23\%$). In contrast, the UC metrics for \emph{University of Aberdeen}, \emph{Loughborough University}, and \emph{University of Brighton} reflect a sustained decline in the same period. Aberdeen and Loughborough both lost significant funding, particularly from EPSRC, which funds comparatively complex research. A comparison of Hull and Brighton reveals that it is the complexity of research undertaken rather than grant income that contributed to the reversal of their relative positions in the ranking. 

To better investigate dynamics of university competitiveness, we construct a dis-aggregated UC$_{u,f,p}$ for three periods at the level of research councils, where $_f$ indicates funder. We compare rankings for two consecutive 5-y periods. Fig.~\ref{fig:evol_uni}b reveals three principal facts: i) across all research councils, the university rankings changed more around the 2015 election than  around the 2010 election, namely the Kendall's $\tau$ correlation coefficient 
between two successive periods 
decreased at all councils; ii) rank turbulence is weakest at EPSRC (Kendall's $\tau=0.66$ in 2010) and MRC ($\tau=0.68$) and strongest at NERC ($\tau=0.52$), AHRC ($\tau=0.55$) and ESRC ($\tau=0.55$); iii) rank turbulence is apparent all the way through each of these rankings. These findings suggest that grant income is the result of vigorous competition for funding. Competition is strongest in fields with low complexity, i.e. at AHRC and ESRC. Rank turbulence is greater at research councils that received larger funding increases after 2015, i.e. NERC, AHRC and ESRC. Notice that MRC is both the research council with the greatest persistence in grant income rankings and the council funding most research of high complexity. Overall, Fig.~\ref{fig:evol_uni} reveals significant variation in universities' research competitiveness.

\begin{figure*}[ht!]
    \includegraphics[width=16cm]{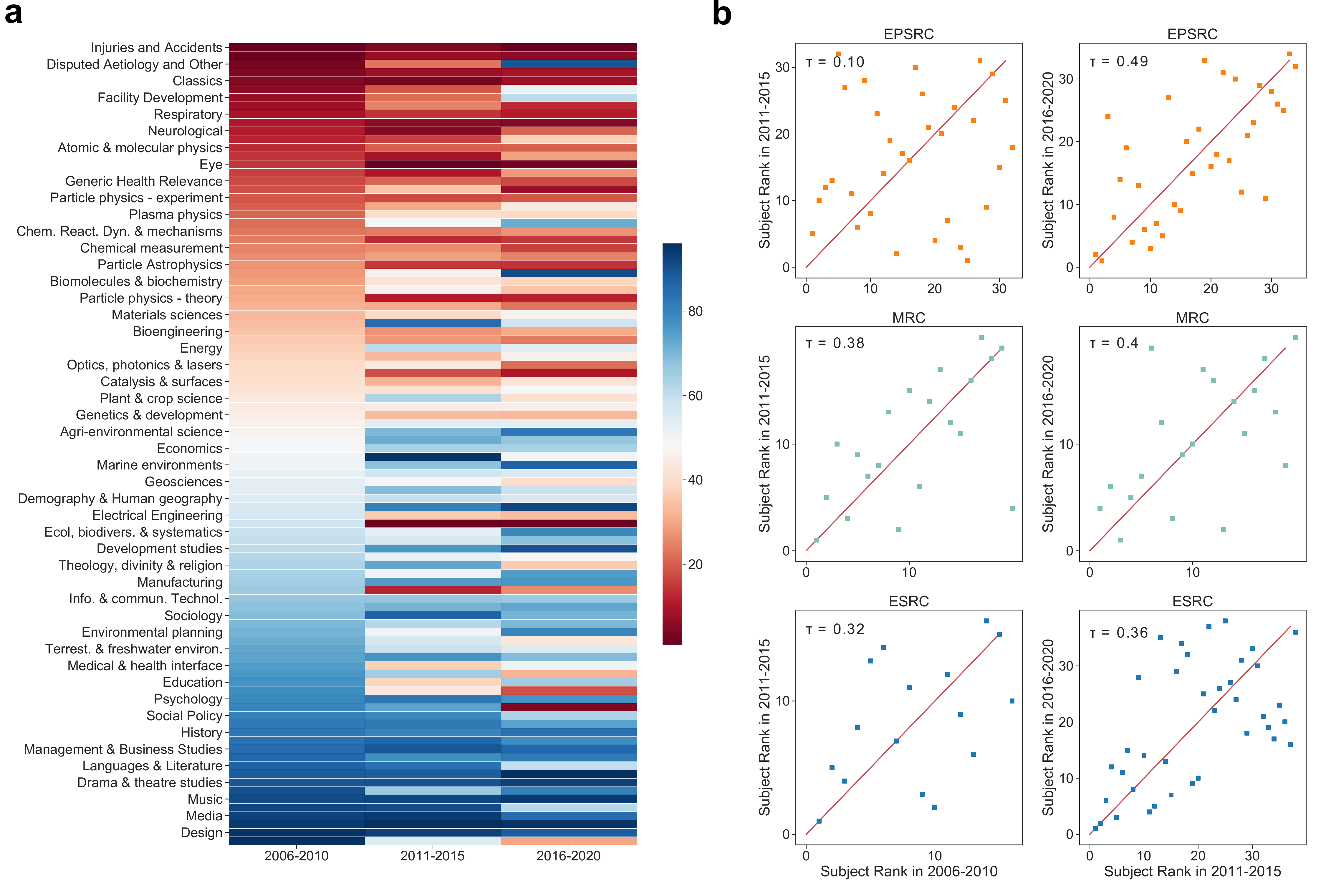}
    \captionsetup{justification=justified}
    \caption{\textbf{Dynamics of subject complexity in the UK.} \textbf{(a)} Dynamics of SC$_{u,p}$ rankings for selected subjects for three 5-y periods, corresponding to three successive UK governments. Full rankings are provided in the \hyperref[sec:suppinf]{SI Appendix}. Research subjects are ranked in descending order by SC$_{u,p}$ for 2006-2010. \textbf{(b)} Dynamics of SC$_{u,f,p}$ rankings at the research council level: each point indicates the ranking of a subject in successive periods. The overall variation in rankings is measured by the reported values of the Kendall’s $\tau$ coefficient.}
    \label{fig:evol_sub}
\end{figure*}

Fig.~\ref{fig:evol_sub}a illustrates the dynamics of research subjects based on SC$_{s,p}$.  Subject complexity exhibits significantly greater variation than university competitiveness. Subjects such as \emph{Environmental Engineering} or \emph{Hearing research} (\emph{Ear}) leaped $60$ places from the bottom to the top of the ranking (Fig.~\ref{fig:evol_sub}a). This is mainly due to universities gaining funding in these subject areas that are already undertaking other complex research. For instance, for the subject of \emph{Ear}, \emph{King's College London} increased its share of funding from $2.1\%$ to $15.7\%$, University of Nottingham from $9.2\%$ to $46.1\%$, while \emph{University of Brighton} started as the largest recipient in this field before $2011$ and currently have no grant income in it at all. We illustrate the changes of subject complexity at the research council level (SC$_{s,f,p}$) by comparing the rankings in two consecutive 5-y periods in Fig.~\ref{fig:evol_sub}b. The subject rankings in the three research councils display similar variability around $2010$ and $2015$. Only at EPSRC variability decreases significantly, with $\tau$ increasing from $0.1$ to $0.49$. This contrasts with increasing turbulence of university rankings (UC$_{u,f,p}$) in this period.

These results on the dynamics of science subjects and universities receiving funding are significant, because they shed light on an important assumption that underpins all efforts to learn about the system of science through metrics: namely that metrics reflect effort and competition rather than links and reputations acquired in the past. We discuss this assumption in the \hyperref[sec:suppinf]{SI Appendix}. Competition for UK grant funding appears to be vigorous in the period we study.

\subsubsection*{Austerity and Grant Income}
\label{sec:ucisci}
\begin{figure*}[ht!]
  \centering
  \begin{subfigure}[t]{0.48\textwidth}
  \begin{subfigure}[t]{0.02\textwidth}
    \textbf{\textsf{a}}
  \end{subfigure}
  \begin{subfigure}[t]{0.9\textwidth}
    \includegraphics[width=\linewidth,valign=t]{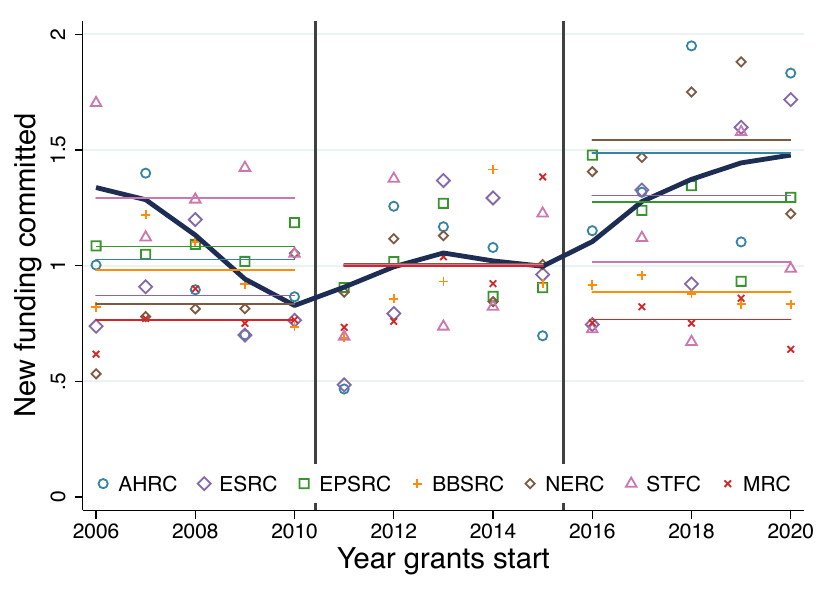}
  \end{subfigure}\hfill \\
  \begin{subfigure}[t]{0.02\textwidth}
    \textbf{\textsf{b}}
  \end{subfigure}
  \begin{subfigure}[t]{0.9\textwidth}
     \includegraphics[width=\linewidth,valign=t]{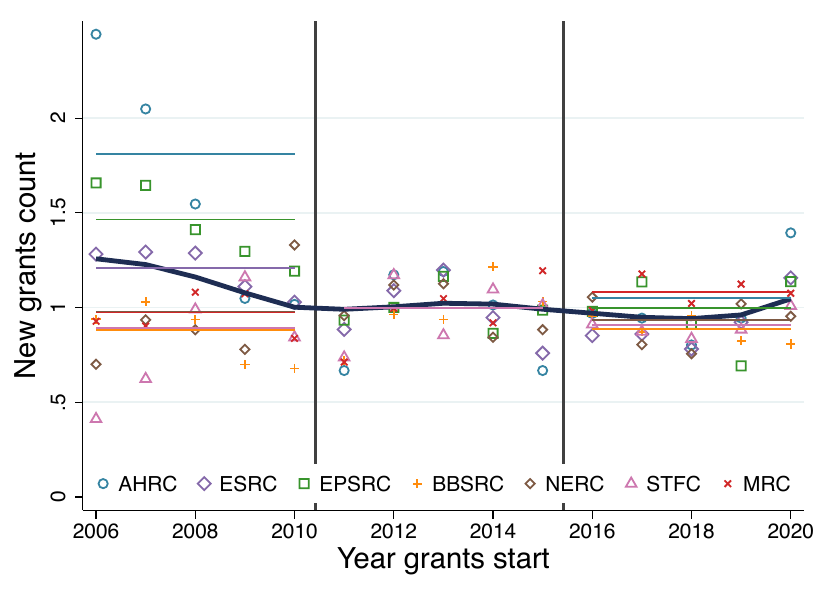}
  \end{subfigure}
  \end{subfigure}
    \begin{subfigure}[t]{0.48\textwidth}
    \begin{subfigure}[t]{0.02\textwidth}
    \textbf{\textsf{c}}
  \end{subfigure}
  \begin{subfigure}[t]{0.9\textwidth}
     \includegraphics[width=\linewidth,valign=t]{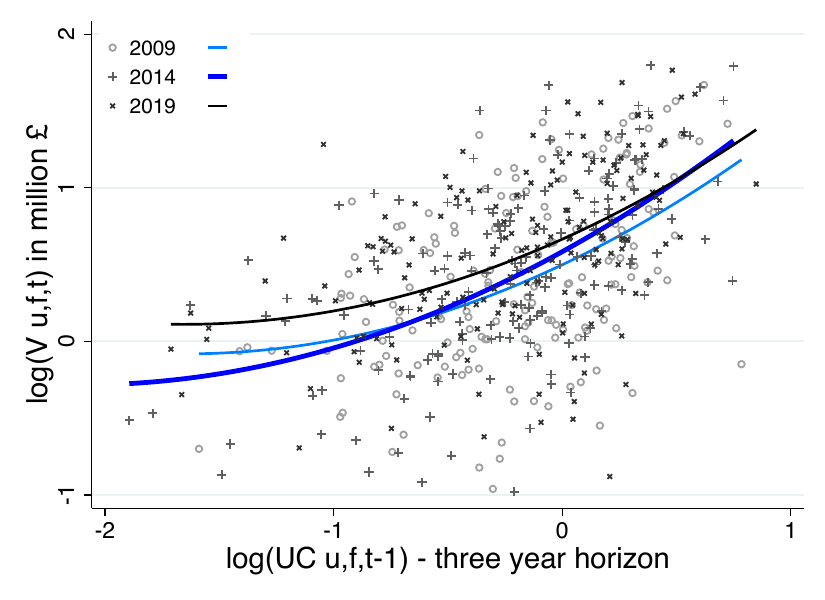}
  \end{subfigure}\hfill \\
  \begin{subfigure}[t]{0.02\textwidth}
    \textbf{\textsf{d}}
  \end{subfigure}
  \begin{subfigure}[t]{0.9\textwidth}
     \includegraphics[width=\linewidth,valign=t]{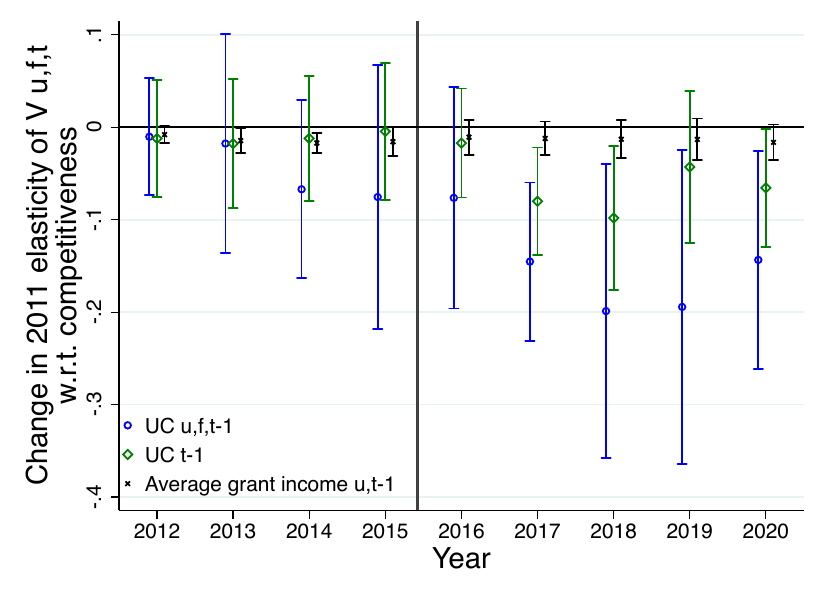}
  \end{subfigure}
  \end{subfigure}
  \captionsetup{justification=justified}
    \caption{ \textbf{Effects of austerity on university grant incomes} $\;$\textbf{(a)} Inflation adjusted new funding for grants relative to 2011-2015. The solid, dark line presents a weighted average of total annual funding relative to 2011-2015. We weight by number of grants each council funds. General elections in May 2010 and May 2015 are indicated by vertical gray bars. $\;$ \textbf{(b)} Count of grants funded relative to 2011-2015. The solid, dark line shows the annual mean for all councils relative to the 2011-2015 average. Total funding was lowest in the years of austerity (2011-2015); grant counts fell before 2011 and then stabilised. $\;$ \textbf{(c)} Scatter of university grant income ($\log$ V$_{u,f,t}$) against $\log$ UC$_{u,f,t-1}$ for the years 2009, 2014 and 2019 with quadratic best fit lines. The correlation of research competitiveness and grant income is highest during austerity. Figure \ref{fig:three_rob}, \hyperref[sec:suppinf]{SI Appendix} explores robustness.  $\;$ \textbf{(d)} Three panel regression estimates of the elasticity of V$_{u,f,t}$ w.r.t. competitiveness (Tables \ref{tab:g}-\ref{tab:gir} \hyperref[sec:suppinf],{SI Appendix}). We contrast coefficients from interacting annual dummies with UC$_{u,f,t}$ (blue circles), UC$_{u,t}$ (green diamonds) and lagged average grant income (black crosses). Exact 95\% confidence intervals are displayed. Solid, vertical, gray line indicates 2015 general election. Where competitiveness is measured using a UC metric the elasticity is constant pre 2016 (parallel trends). Post 2015 it is reduced, confirming the descriptive result shown in panel c: competitiveness counts most during austerity. The elasticity of grant income w.r.t. lagged average grant income is constant post 2015, revealing this is a poor measure of competitiveness.}
    \label{fig:three}
\end{figure*}

In 2010 a coalition between the Liberal Democrats and the Conservative Party in the UK introduced significant restrictions on government spending (austerity) including a freeze of the science budget in nominal terms \cite{LGS18,JSS19}. 
Surprisingly, the Conservative Party were returned to power in the 2015 UK general election \cite{GP16}. The new government announced that science funding would be adjusted for inflation, relaxing austerity for science funding. The political uncertainty before the election suggests no scientists could anticipate how science funding would develop after 2015. This provides a natural experiment. We use it to analyse how research competitiveness affected university grant income. 

Drawing on economic models of contests \cite{MS06,FNS20,OS20} we expect that austerity increased the importance of research competitiveness in determining universities' grant incomes. Ref.~\cite{MS06} shows that reducing the number of prizes (grants) increases efforts of high ability contestants, but decreases efforts of low ability contestants. This prediction requires costs of participating in the contest are convex in effort, which does not seem unwarranted in this context. In the UK the number of grants offered across all councils fell by 20\% after 2008 (Fig. \ref{fig:three}b). At the same time total funds committed to grants fell reaching a minimum in 2010. Fig. \ref{fig:three}c confirms the correlation of grant income and UC$_{u,f,t-1}$ increased during austerity. Fig. \ref{fig:three}a shows that in terms of funding committed to grants austerity was relaxed from 2015. At this time the correlation between grant income and competitiveness (UC$_{u,f,t-1}$) shown in Fig.~\ref{fig:three}c decreases.

To test whether research competitiveness as measured by UC$_{u,f,t}$ had a causal effect on grant incomes we estimate difference-in-differences models \cite{IW09,AP10}, focusing on the years 2011-2020. We estimate a generalized model with continuous treatment intensity (UC$_{u,f,t-1}$) \cite{AAL04} and use Poisson regression models to retain years in which universities have no grant income (\hyperref[sec:m]{Methods}). Results confirm that post 2015 the slope of the relationship between research competitiveness and grant income is flatter in a range of models that underscore robustness of our findings (\hyperref[sec:suppinf]{SI Appendix}). A widely used test of the parallel trends assumption underpinning difference-in-differences models is estimated using a flexible functional form \cite{dhA03}. Fig.~\ref{fig:three}d presents coefficients and exact 95\% confidence intervals from interacting two UC measures with annual dummies. The coefficients for UC$_{u,t-1}$ and UC$_{u,f,t-1}$ turn negative after 2015, as expected. Fig.~\ref{fig:three}d also presents coefficients and exact 95\% confidence intervals for average grant income as an alternative and much simpler measure of university competitiveness. These coefficients become statistically significant before 2015 and their profile does not indicate that the effect of lagged grant income on current grant income changed after 2015. These results show that average grant income is a poor measure of research competitiveness: it does not capture the causal effect of research competitiveness on grant income in a way theory predicts. This failure of the lagged grant income measure is also reflected in Fig.~\ref{fig:fs_rob}d \hyperref[sec:suppinf]({SI Appendix}), which shows a falling correlation between grant incomes and lagged average grant income over time. That drop in correlation is reflected in our finding above that university rankings became more turbulent after 2010.

 Our estimates imply that a one standard deviation increase in research competitiveness at the mean would have resulted in a grant income increase of 10\% prior to 2015, around \pounds 240,000 per funder and year. After 2015 this falls to 6.6\%, illustrating the loss of advantage experienced by the more research competitive universities at this time.

In \hyperref[sec:m]{Methods} we discuss robustness tests. Details are set out in the \hyperref[sec:suppinf]{SI Appendix}. All regressions reported there use time-series of UC metrics. 

\section*{Discussion}

University rankings are increasingly important: they act as signposts for students and employers, attract donors and young scientists. Existing rankings are frequently compared and studied~\cite{hfM17,VBM18}. University leaders recognise that the perception of excellence drives competition for talent, intensifying  pressure to move up the rankings. Given the influence of rankings, these ought to derive from solid evidence and a clear methodology. It is widely recognised that this is not the case~\cite{yG16,BL20} and that the impact of rankings creates pressures to manipulate input data~\cite{mS96}. Rankings designed with these challenges in mind are sorely needed~\cite{jW15,WB20}.

University rankings also matter for the evaluation of science funding. Any attempt to understand how universities adjust to changes in science funding or changes in opportunities for basic research requires metrics that allow ranking of universities. Currently most rankings rely on a mix of bibliometric and grant income data; typically the resulting measures are weighted aggregates of the input data and lag current university activity.

In this paper we have proposed a method for ranking university research competitiveness that is based on up to date information for university research activities, that is robust to the dynamics of science and that is difficult to manipulate. Most, if not all, university rankings currently available do not reach this standard. Applying our method to study the grant funding in the UK, we find that research competitiveness has a stronger effect on grant income in periods of austerity. We also show that increases in the amount of funding available tend to percolate down to less competitive research groups. More work is needed to determine the welfare implications of such funding increases.

Much more work needs to be done to link grant income to research outputs and measures of impact. Once these data have been constructed it will be possible to develop insights into the welfare effects of different ways of distributing grant funding to universities using this or similar measures of research competitiveness.

More broadly previous work on complexity measures ~\cite{HH09, cH21} has used these in panel regressions. We are not aware of work employing time series of complexity measures to study effects of a natural experiment. We have explored a range of alternative approaches to constructing measures of university competitiveness drawing on the literature \cite{cH21}. Our main results are robust to changes in how the proposed measure is constructed. More work can be done to establish the properties of this and similar measures of complexity/competitiveness in the context of panel data regressions. 

\section*{Methods}
\label{sec:m}

\subsection*{Data sets} We used the research grants data collected from UK Research and Innovation (UKRI), a funding body overseeing all UK research councils and Research England. The dataset covers $43,430$ research grants from seven national research councils conducted between $2006$ and $2020$ (Table~\ref{tab:grant}). Each record contains information on the lead universities, investigators (principal investigators and co-investigators), grant value and the percentage composition of research subject areas within the grant. Grants recorded comprise a broad spectrum of academic disciplines including medical and biological sciences, astronomy, physics, chemistry and engineering, social sciences, economics, environmental sciences and the arts and humanities, which allows us to comprehensively investigate the research and innovation in the UK. Note that here grants are considered to have been awarded to the lead university, and allocated to each of the listed research subjects according to percentage shares recorded in the grant data. 

\subsection*{University-subject bipartite matrix}
To measure the relative importance of different universities in the same research subject, we represent grant income data as a weighted bipartite matrix, whose nodes are of two types: universities and research subjects. Links can only exist between the universities and research subjects. An element $M_{u,s}$ in the weighted bipartite matrix is defined as:
\begin{equation}
M_{u,s} = \frac{V_{u,s}}{\sum_{u}V_{u,s}}
\end{equation}
where $V_{u,s}$ is the amount of funding received by university $u$ in the research subject $s$. $M_{u,s}$ represents the fraction of funding volume in subject $s$ awarded to the university $u$, So that we have $\sum_{u}M_{u,s}=1$ for each subject $s$. By this normalization, all the research subjects are placed on equal status, regardless of the difference in awarded funding amount. For the sake of robustness, when building the bipartite network at the overall level, we have considered $77$ universities that received at least one research grant per year, and $101$ research subjects that appeared at least nine times over the whole period. While at the research council level, we have selected the universities that received grants at least in $5$ subjects, and the subjects that were conducted by at least $5$ universities within each time window.

\subsection*{Measuring University Competitiveness and Subject Complexity} Inspired by the high level of nestedness~\cite{mariani2019nestedness} observed in our university-subject network (as be seen in Fig.~\ref{fig:overall_feature}), we employ the non-linear iterative algorithm of Tacchella et al.~\cite{tacchella2012new} to quantify the University Competitiveness (UC) and the Subject Complexity (SC) metrics. The algorithm assumes that the competitiveness of a university is determined by the proportional sum of the subjects weighted by their complexity. 
The larger is the number of research subjects in which they have obtained funding, and the more sophisticated these subjects are, the more competitive the university is. On the other hand, the complexity of a subject is assumed to be inversely proportional to the number of universities which have received funding, and mainly determined by the competitiveness of the less competitive universities active in such a subject. 

The iterative algorithm can be expressed as in Eq.~\ref{eq:NonlinearAlgorithm}, where we denote the competitiveness of university $u$ as UC$_{u}$ and the complexity of subject $s$ as SC$_{s}$:

\begin{align}
\left\{
\begin{array}{rcl}
\widetilde{UC}^{(n)}_{u} &=\sum_{s}M_{us}SC^{(n-1)}_{s}, \\
\widetilde{SC}^{(n)}_{s} &=\frac{1}{\sum_{u}M_{us}\frac{1}{UC^{(n-1)}_{u}}}
\end{array} \right. \rightarrow
\left\{
\begin{array}{rcl}
 UC^{(n)}_{u} = \frac{\widetilde{UC}^{(n)}_{u}}{\left \langle \widetilde{UC}^{(n)}_{u} \right \rangle}, \\
SC^{(n)}_{s} = \frac{\widetilde{SC}^{(n)}_{s}}{\left \langle\widetilde{SC}^{(n)}_{s}\right \rangle}
\end{array} \right.
\label{eq:NonlinearAlgorithm}
\end{align}

The algorithm starts from even values with ${UC}_{u}^{(0)}=1 \forall u$ and ${SC}_{s}^{(0)}=1 \forall s$. At each iteration, we compute the intermediate variables $\widetilde{UC}^{(n)}_{u}$ and $\widetilde{SC}^{(n)}_{s}$ and then normalize them. The values of university competitiveness and subject complexity converge to a fix point. The resulting values are used to rank universities and research subjects. Note that although the algorithm can be applied to any bipartite network, the convergence properties of the algorithm are determined by the structure of the bipartite matrix. Therefore, to ensure that the measures of competitiveness and complexity are reliable, we have tested the fixed points of the algorithm implemented on all the aggregated bipartite networks considered in this study, finding the these quantities all converge to non-zero values that are independent of the initial conditions.

\subsection*{Regression Models Analysing Austerity} To establish how research competitiveness affected university grant incomes after austerity ended we estimate difference-in-differences models using panel data for the years 2011-2020. The 2015 general election in the UK resulted in a significant and unanticipated change to the trajectory of grant funding, due to the end of strict austerity for research councils \cite{GP16}. This provides an opportunity to test whether variation in university competitiveness in 2015 is a significant predictor of grant incomes in subsequent years. We include university and research council fixed effects as well as start-year fixed effects for grant income at the university-research council level in all models reported in the \hyperref[sec:suppinf]{SI Appendix} and Fig. \ref{fig:three}c. Many universities do not generate grant income (V$_{u,f,t}$) in every year, therefore we treat grant income as a count variable and estimate fixed-effects Poisson models \cite{ST06,CF18}:

\begin{align}
\label{eq:model}
    \text{V}_{u,f,t}=&\exp(\alpha \text{D}_{Post} +\delta \text{D}_{Post} \times \text{UC}_{u,f,t-1} \notag \\ &\qquad\qquad\qquad +\text{X}'_{u,f,t}\beta+\lambda_t+\gamma_{u,f})+\epsilon_{u,f,t}\quad .
\end{align}

In these models the coefficient ($\delta$) on the interaction of lagged UC$_{u,f,t-1}$ and the post 2015 dummy variable ($D_{Post}$) captures the causal effect of university competitiveness on grant income in subsequent years, if there are no time varying unobserved changes in universities' reputations or unobserved changes in the strength of unobserved links between universities and research councils that affect ability to obtain grants around 2015 \cite{AP10,WKBG18,sC21}. We test the key identifying assumption of parallel trends by estimating a more flexible version of the above specification in which we interact UC$_{u,f,t-1}$ with annual dummies \cite{dhA03}. Coefficients are reported in Fig.\ref{fig:three}c, which shows that the elasticity of grant incomes w.r.t research competitiveness ($\delta_t$) fell when austerity was relaxed in 2015. Importantly, this elasticity did not change in the years prior to 2015. This finding indicates that the elasticity remained constant before austerity was relaxed and the change in the elasticity resulted from the policy shock of 2015. Note that all models include covariates ($X_{u,f,t}$) capturing changes in research council policies: a count of grants awarded in the previous year (NG$_{u,f,t-1}$), the median grant income in the previous year (md V$_{f,t-1}$) and total grant expenditure in the previous year ($\Sigma \text{V}_{f,t-1}$). Models also include a proxy for university reputation (R$_{u,f,t-1}$) measured by the count of grants awarded by all other research councils in the previous year. Lastly, several models include the Herfindahl index (HHI$_{f,t}$) measuring concentration of grant income at the research council. This is added to exclude that the UC metric captures grant income concentration, an interpretation that has been advanced and rejected in other contexts \cite{cH21}.

Specifications that test robustness of our findings are reported in the \hyperref[sec:suppinf]{SI Appendix}.  We test a further variant of the UC metric and a simpler competitiveness measure using average grant income in the three previous years ($\widebar{\text{V}}_{u,f,t-1}$). Fig.\ref{fig:three}c shows this simpler measure of university competitiveness fails to detect the relaxation of austerity correctly. 

As noted, the end of strict austerity was announced after a surprising election outcome \cite{GP16}. This rules out that applicants or research councils anticipated more funding would become available after the 2015  election, undermining strict exogeneity. Recent work lists robustness tests that further support a causal interpretation of the results obtained from difference in differences models \cite{WKBG18,sC21}. Our main data do not include information on the number of grant applications that failed, but we have obtained such data for the years 2015-2019. This allows us to examine differences in the response to the 2015 shock by research competitiveness. Figure \ref{fig:fs_rob} in the \hyperref[sec:suppinf]{SI Appendix} reveals that after 2015 universities increased the number of grant applications and total value of grant income requested more, if they were less research competitive. These findings are in line with the predictions of contest theory \cite{MS06,FNS20,OS20}. To further test our results we vary the specification estimated. Tables in the \hyperref[sec:suppinf]{SI Appendix} set out results from models in which we: i) include research council time trends, ii) reduce the number of years prior to the policy shock included in the models and iii) use OLS instead of Poisson models. All of these tests support our main findings.   

\newpage

\bibliography{sr-sample}

\begin{thebibliography}{10}

\bibitem{BS15}
C.~Bloch and M.~P. S{\o}rensen, ``The size of research funding: Trends and
  implications,'' {\em Science and Public Policy}, vol.~42, no.~1, pp.~30--43,
  2015.

\bibitem{MML15}
A.~Ma, R.~J. Mondrag{\'o}n, and V.~Latora, ``Anatomy of funded research in
  science,'' {\em Proceedings of the National Academy of Sciences}, vol.~112,
  no.~48, pp.~14760--14765, 2015.

\bibitem{NA21}
M.~W. Nielsen and J.~P. Andersen, ``Global citation inequality is on the
  rise,'' {\em Proceedings of the National Academy of Sciences}, vol.~118,
  no.~7, 2021.

\bibitem{wpW19}
W.~P. Wahls, ``Opinion\protect{:} the \protect{National Institutes of Health}
  needs to better balance funding distributions among \protect{US}
  institutions,'' {\em Proceedings of the National Academy of Sciences},
  vol.~116, no.~27, pp.~13150--13154, 2019.

\bibitem{bJ09}
B.~F. Jones, ``The burden of knowledge and the “\protect{Death} of the
  \protect{Renaissance Man}”: Is innovation getting harder?,'' {\em The
  Review of Economic Studies}, vol.~76, no.~1, pp.~283--317, 2009.

\bibitem{ABD20}
T.~Astebro, S.~Braguinsky, and Y.~Ding, ``Declining business dynamism among our
  best opportunities: The role of the burden of knowledge,'' Working Paper
  27787, National Bureau of Economic Research, September 2020.

\bibitem{LGS18}
P.~Larrue, D.~Guellec, and F.~Sgard, {\em OECD Science, Technology and
  Innovation Outlook 2018}, ch.~New trends in public research funding,
  pp.~185--204.
\newblock OECD, 2018.

\bibitem{JSS19}
J.~Janger, N.~Schmidt, and A.~Strauss, ``International differences in basic
  research grant funding-a systematic comparison,'' tech. rep., Studien zum
  deutschen Innovationssystem, 2019.

\bibitem{rkM68}
R.~K. Merton, ``The \protect{Matthew} effect in science: The reward and
  communication systems of science are considered,'' {\em Science}, vol.~159,
  no.~3810, pp.~56--63, 1968.

\bibitem{ASW14}
P.~Azoulay, T.~Stuart, and Y.~Wang, ``Matthew: Effect or fable?,'' {\em
  Management Science}, vol.~60, no.~1, pp.~92--109, 2014.

\bibitem{BVR18}
T.~Bol, M.~de~Vaan, and A.~van~de Rijt, ``The \protect{Matthew} effect in
  science funding,'' {\em Proceedings of the National Academy of Sciences},
  vol.~115, no.~19, pp.~4887--4890, 2018.

\bibitem{MS06}
B.~Moldovanu and A.~Sela, ``Contest architecture,'' {\em Journal of Economic
  Theory}, vol.~126, no.~1, pp.~70--96, 2006.

\bibitem{FNS20}
D.~Fang, T.~Noe, and P.~Strack, ``Turning up the heat: The discouraging effect
  of competition in contests,'' {\em Journal of Political Economy}, vol.~128,
  no.~5, pp.~1940--1975, 2020.

\bibitem{OS20}
W.~Olszewski and R.~Siegel, ``Performance-maximizing large contests,'' {\em
  Theoretical Economics}, vol.~15, no.~1, pp.~57--88, 2020.

\bibitem{GP16}
J.~Green and C.~Prosser, ``Party system fragmentation and single-party
  government: the british general election of 2015,'' {\em West European
  Politics}, vol.~39, no.~6, pp.~1299--1310, 2016.

\bibitem{JJ18}
J.~Johnes, ``University rankings: What do they really show?,'' {\em
  Scientometrics}, vol.~115, no.~1, pp.~585--606, 2018.

\bibitem{VBM18}
M.~M. Vernon, E.~A. Balas, and S.~Momani, ``Are university rankings useful to
  improve research? a systematic review,'' {\em PloS One}, vol.~13, no.~3,
  p.~e0193762, 2018.

\bibitem{GWB19}
S.~Gralka, K.~Wohlrabe, and L.~Bornmann, ``How to measure research efficiency
  in higher education? research grants vs. publication output,'' {\em Journal
  of Higher Education Policy and Management}, vol.~41, no.~3, pp.~322--341,
  2019.

\bibitem{dora}
``Declaration on research assessment,'' May 2013.
\newblock [Online; posted May 13, 2013].

\bibitem{LM15}
D.~Hicks, P.~Wouters, L.~Waltman, S.~De~Rijcke, and I.~Rafols, ``Bibliometrics:
  the leiden manifesto for research metrics,'' {\em Nature News}, vol.~520,
  no.~7548, p.~429, 2015.

\bibitem{dec}
``Statement by three national academies,'' October 2017.
\newblock [Online; posted October 27, 2017].

\bibitem{MB04}
K.~K. Mane and K.~B{\"o}rner, ``Mapping topics and topic bursts in
  \protect{PNAS},'' {\em Proceedings of the National Academy of Sciences},
  vol.~101, no.~suppl 1, pp.~5287--5290, 2004.

\bibitem{KPH14}
T.~Kuhn, M.~c.~v. Perc, and D.~Helbing, ``Inheritance patterns in citation
  networks reveal scientific memes,'' {\em Phys. Rev. X}, vol.~4, p.~041036,
  Nov 2014.

\bibitem{HH09}
C.~A. Hidalgo and R.~Hausmann, ``The building blocks of economic complexity,''
  {\em Proceedings of the National Academy of Sciences}, vol.~106, no.~26,
  pp.~10570--10575, 2009.

\bibitem{cH21}
C.~A. Hidalgo, ``Economic complexity theory and applications,'' {\em Nature
  Reviews Physics}, pp.~1--22, 2021.

\bibitem{tacchella2012new}
A.~Tacchella, M.~Cristelli, G.~Caldarelli, A.~Gabrielli, and L.~Pietronero, ``A
  new metrics for countries' fitness and products' complexity,'' {\em Sci.
  Rep.}, vol.~2, p.~723, 2012.

\bibitem{MFT19}
P.~Mealy, J.~D. Farmer, and A.~Teytelboym, ``Interpreting economic
  complexity,'' {\em Science Advances}, vol.~5, no.~1, p.~eaau1705, 2019.

\bibitem{FM21}
B.~S. Fritz and R.~A. Manduca, ``The economic complexity of \protect{US}
  metropolitan areas,'' {\em Regional Studies}, pp.~1--12, 2021.

\bibitem{hfM17}
H.~F. Moed, ``A critical comparative analysis of five world university
  rankings,'' {\em Scientometrics}, vol.~110, no.~2, pp.~967--990, 2017.

\bibitem{DGSA18}
L.~Dias, M.~Gerlach, J.~Scharloth, and E.~G. Altmann, ``Using text analysis to
  quantify the similarity and evolution of scientific disciplines,'' {\em Royal
  Society Open Science}, vol.~5, no.~1, pp.~171--545, 2018.

\bibitem{SBWTS21}
C.~Singh, E.~Barme, R.~Ward, L.~Tupikina, and M.~Santolini, ``Quantifying the
  rise and fall of scientific fields,'' {\em arXiv preprint arXiv:2107.03749},
  2021.

\bibitem{IW09}
G.~W. Imbens and J.~M. Wooldridge, ``Recent developments in the econometrics of
  program evaluation,'' {\em Journal of Economic Literature}, vol.~47, no.~1,
  pp.~5--86, 2009.

\bibitem{AP10}
J.~D. Angrist and J.-S. Pischke, ``The credibility revolution in empirical
  economics: How better research design is taking the con out of
  econometrics,'' {\em Journal of Economic Perspectives}, vol.~24, no.~2,
  pp.~3--30, 2010.

\bibitem{AAL04}
D.~Acemoglu, D.~H. Autor, and D.~Lyle, ``Women, war, and wages: The effect of
  female labor supply on the wage structure at midcentury,'' {\em Journal of
  Political Economy}, vol.~112, no.~3, pp.~497--551, 2004.

\bibitem{dhA03}
D.~Autor, ``Outsourcing at will: The contribution of unjust dismissal doctrine
  to the growth of employment outsourcing,'' {\em Journal of Labor Economics},
  vol.~21, no.~1, pp.~1--42, 2003.

\bibitem{yG16}
Y.~Gingras, {\em Bibliometrics and Research Evaluation: Uses and Abuses}.
\newblock MIT Press, 2016.

\bibitem{BL20}
M.~Biagioli and A.~Lippman, eds., {\em Gaming the Metrics: Misconduct and
  Manipulation in Academic Research}.
\newblock MIT Press, 2020.

\bibitem{mS96}
M.~Strathern, ``From improvement to enhancement: An anthropological comment on
  the audit culture,'' {\em Cambridge Anthropology}, vol.~19, no.~3, pp.~1--21,
  1996.

\bibitem{jW15}
J.~Wilsdon, L.~Allen, E.~Belfiore, P.~Campbell, S.~Curry, S.~Hill, R.~Jones,
  R.~Kain, S.~Kerridge, M.~Thelwall, {\em et~al.}, ``The metric tide,'' tech.
  rep., Report of the Independent Review of the Role of Metrics in Research
  Assessment and Management, 2015.

\bibitem{WB20}
J.~D. West and C.~T. Bergstrom, ``Misinformation in and about science,'' {\em
  Proceedings of the National Academy of Sciences}, vol.~118, no.~15, 2021.

\bibitem{mariani2019nestedness}
M.~S. Mariani, Z.-M. Ren, J.~Bascompte, and C.~J. Tessone, ``Nestedness in
  complex networks: observation, emergence, and implications,'' {\em Phys.
  Rep.}, vol.~813, pp.~1--90, 2019.

\bibitem{ST06}
J.~S. Silva and S.~Tenreyro, ``The log of gravity,'' {\em The Review of
  Economics and Statistics}, vol.~88, no.~4, pp.~641--658, 2006.

\bibitem{CF18}
E.~Ciani and P.~Fisher, ``Dif-in-dif estimators of multiplicative treatment
  effects,'' {\em Journal of Econometric Methods}, vol.~8, no.~1, 2018.

\bibitem{WKBG18}
C.~Wing, K.~Simon, and R.~A. Bello-Gomez, ``Designing difference in difference
  studies: best practices for public health policy research,'' {\em Annual
  Review of Public Health}, vol.~39, 2018.

\bibitem{sC21}
S.~Cunningham, {\em Causal Inference: The Mixtape}.
\newblock Yale University Press, 2021.

\bibitem{mB18}
M.~Baldwin, ``Scientific autonomy, public accountability, and the rise of
  “peer review” in the cold war \protect{United States},'' {\em Isis},
  vol.~109, no.~3, pp.~538--558, 2018.

\bibitem{AL20}
P.~Azoulay and D.~Li, ``Scientific grant funding,'' Working Paper 26889,
  National Bureau of Economic Research, March 2020.

\bibitem{bG06}
B.~Godin, ``On the origins of bibliometrics,'' {\em Scientometrics}, vol.~68,
  no.~1, pp.~109--133, 2006.

\bibitem{aC20}
A.~Csiszar, {\em Gaming the Metrics}, ch.~1. Gaming Metrics Before the Game,
  pp.~26--31.
\newblock Cambridge, Mass.: MIT Press, 2020.

\bibitem{ajP89}
A.~Phillimore, ``University research performance indicators in practice: The
  university grants committee's evaluation of british universities, 1985--86,''
  {\em Research Policy}, vol.~18, no.~5, pp.~255--271, 1989.

\bibitem{eG06}
E.~Garfield, ``The history and meaning of the journal impact factor,'' {\em
  JAMA}, vol.~295, no.~1, pp.~90--93, 2006.

\bibitem{bK20}
B.~Kehm, {\em Gaming the Metrics}, ch.~6. Global University Rankings: Impacts
  and Applications, pp.~93--100.
\newblock Cambridge, Mass.: MIT Press, 2020.

\bibitem{yG20}
Y.~Gingras, {\em Gaming the Metrics}, ch.~2. The transformation of the
  Scientific Paper, pp.~32--43.
\newblock Cambridge, Mass.: MIT Press, 2020.

\bibitem{iR18}
I.~Rowlands, ``{What are we measuring? Refocusing on some fundamentals in the
  age of desktop bibliometrics},'' {\em FEMS Microbiology Letters}, vol.~365,
  03 2018.

\bibitem{N15}
P.~Nurse, ``Ensuring a successful \protect{UK} research endeavour,'' tech.
  rep., Department for Business, Innovation and Skills, 2015.

\bibitem{gS17}
G.~Sivertsen, ``Unique, but still best practice? the research excellence
  framework (ref) from an international perspective,'' {\em Palgrave
  Communications}, vol.~3, no.~1, pp.~1--6, 2017.

\end{thebibliography}

\newpage

\appendix
\label{sec:suppinf}

\section*{Supporting Information}
This appendix provides detail on several figures (\ref{sec:si_desfig}), an extensive discussion of econometric results and robustness checks (\ref{sec:si_eres}) not included in the main paper and a model of peer review that motivates our empirical model  (\ref{app:fsci}).

\section{Detailed Figures and Tables}
\label{sec:si_desfig}
\begin{figure*}[ht!]
\centering
    \includegraphics[width=17cm]{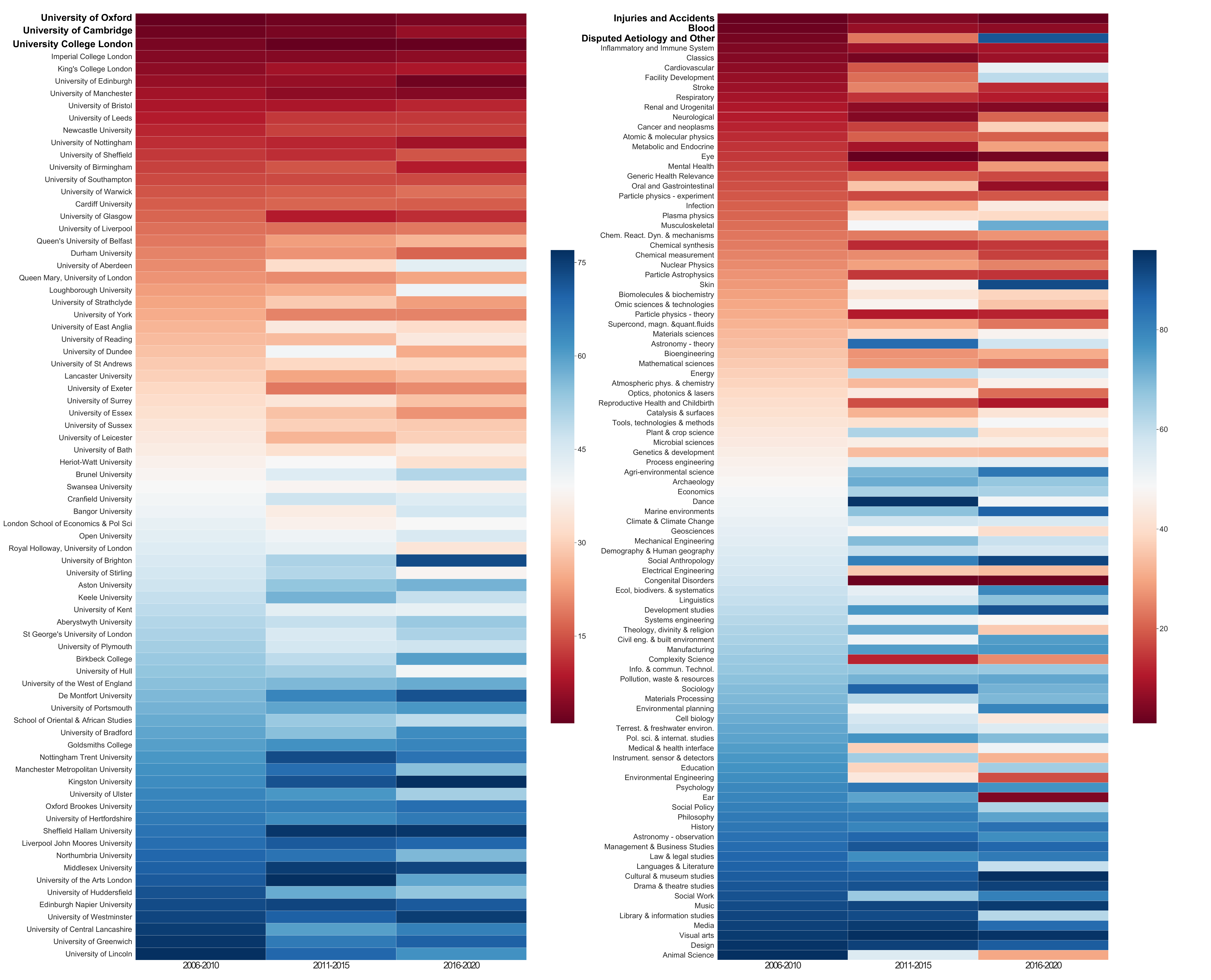}
    \captionsetup{justification=justified}
    \caption{ \textbf{Evolution of university competitiveness and subject complexity over three 5-y periods.} All universities (left panel) and research subjects (right panel) are sorted in descending order according to their rank in the period $2006-2010$.}
    \label{fig:Evolution_University_Subject}
\end{figure*}

\begin{figure*}[ht!]
\centering
    \includegraphics[width=14cm]{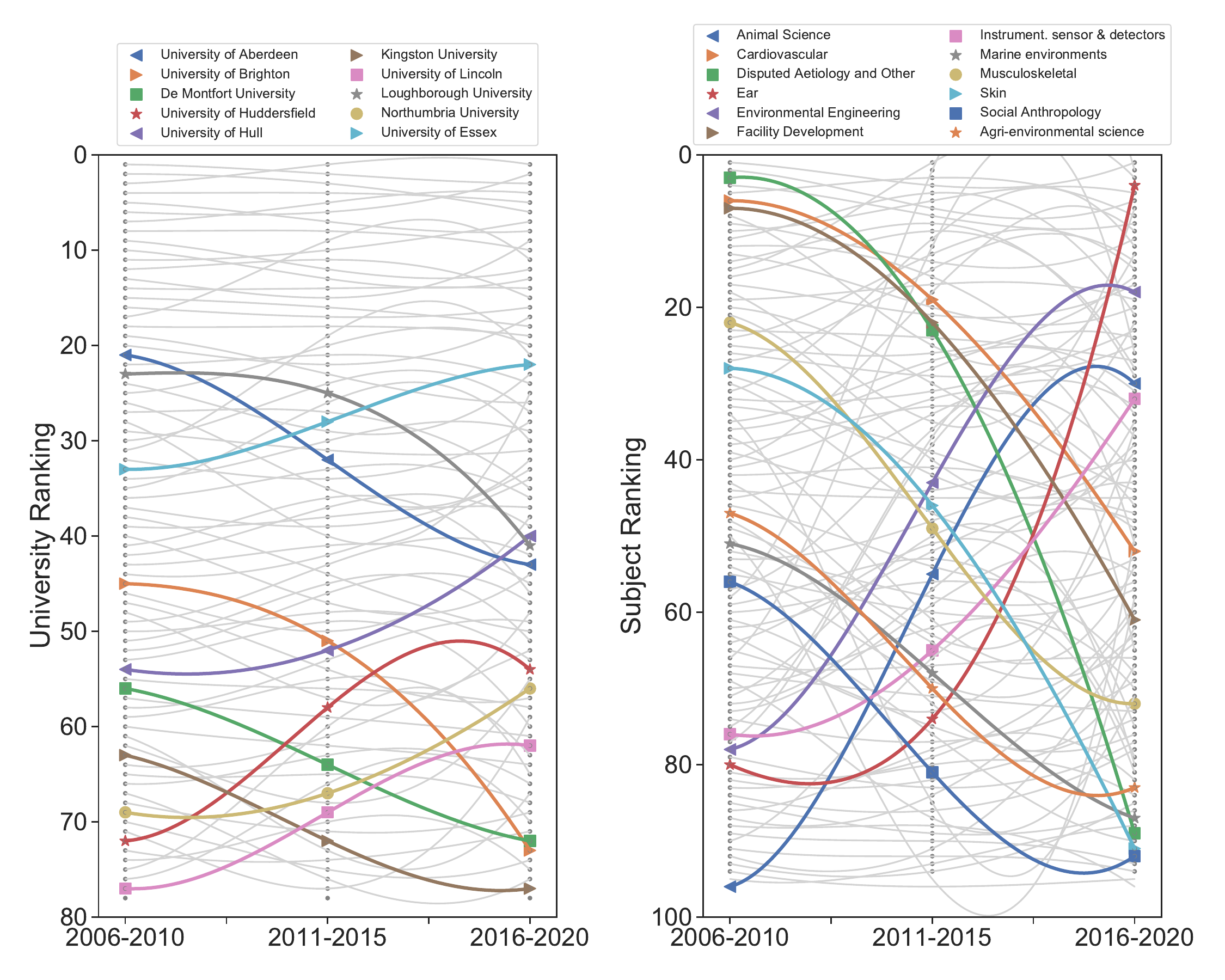}
    \captionsetup{justification=justified}
    \caption{\textbf{Evolution of $10$ selected universities and $12$ selected subjects}. Selected universities and subjects have the most significant changes in competitiveness and complexity rankings over three 5-year time windows. Universities and subjects are sorted in descending order according their average rankings of competitiveness and complexity, respectively. The evolution of other universities and subjects are shown in grey.}
    \label{fig:Evolution_Uni_Sub_Emphasis}
\end{figure*}

Fig. \ref{fig:Evolution_University_Subject} shows the dynamics of rankings for all universities and subjects contained in our data. The rankings of universities and subjects shown in Fig. \ref{fig:evol_uni} and \ref{fig:evol_sub} are based on subsets drawn from the rankings shown here. Additional detail on the dynamics of selected universities and subjects is given in Fig. \ref{fig:Evolution_Uni_Sub_Emphasis}. The figure highlights universities and subjects experiencing the largest change in their ranks between the first and last period.

\begin{figure*}[hb!]
  \centering
  \begin{subfigure}[t]{0.02\textwidth}
    \textbf{\textsf{a}}
  \end{subfigure}
  \begin{subfigure}[t]{0.3\textwidth}
    \includegraphics[width=\linewidth,valign=t]{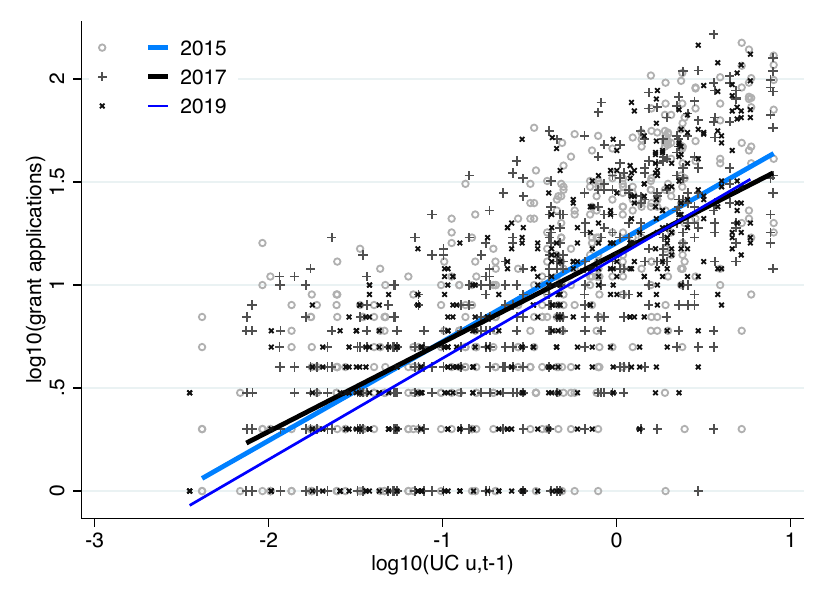}
  \end{subfigure}
  \begin{subfigure}[t]{0.02\textwidth}
    \textbf{\textsf{b}}
  \end{subfigure}
  \begin{subfigure}[t]{0.3\textwidth}
     \includegraphics[width=\linewidth,valign=t]{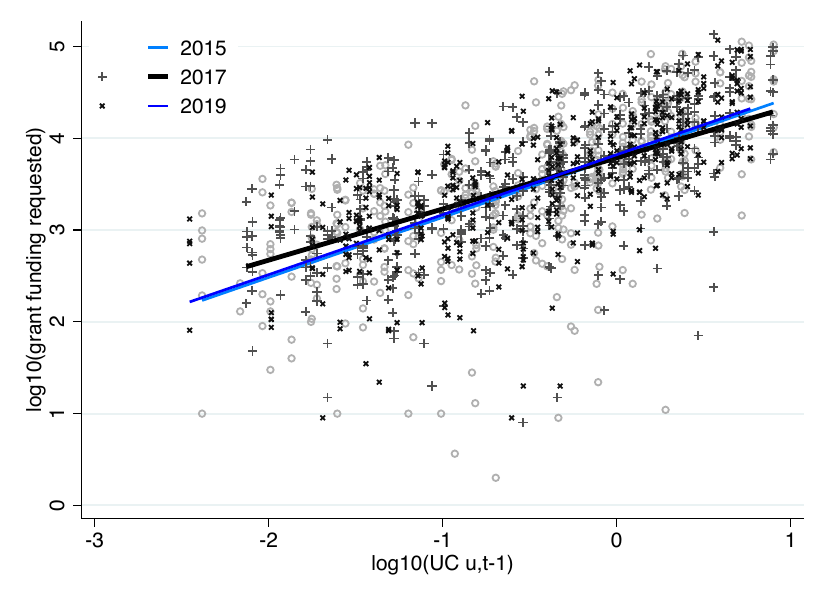}
  \end{subfigure}
  \begin{subfigure}[t]{0.02\textwidth}
    \textbf{\textsf{c}}
  \end{subfigure}
  \begin{subfigure}[t]{0.3\textwidth}
     \includegraphics[width=\linewidth,valign=t]{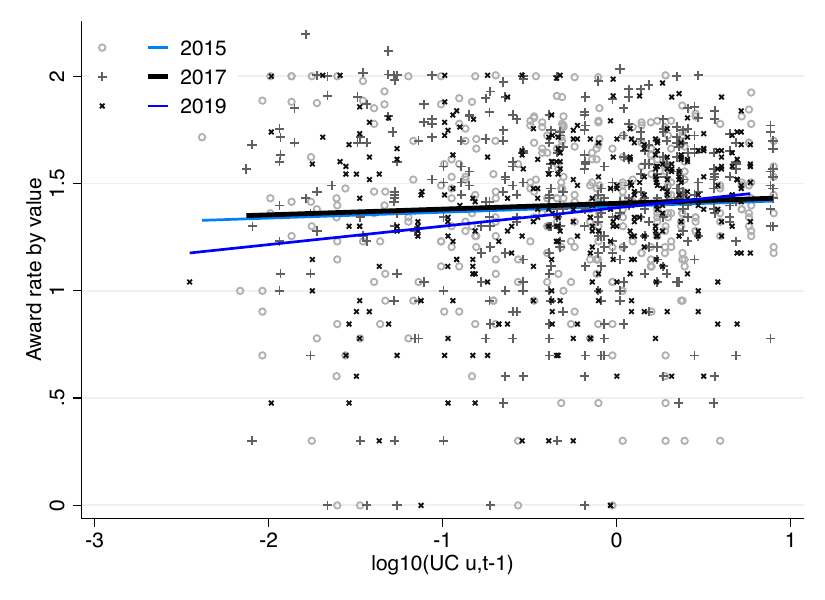}
  \end{subfigure}
    \captionsetup{justification=justified}
    \caption{\textbf{Applications and outcomes 2015-2019}. \textbf{(a)} shows that the elasticity of application counts w.r.t. UC$_{u,t}$ fell in 2017 relative to 2015. \textbf{(b)} shows that the elasticity of application values w.r.t. UC$_{u,t}$ also fell in 2017 relative to 2015. Relative to 2015 the number and value of applications was greater in 2017 for the universities with lowest research competitiveness and lower in 2017 for universities with highest research competitiveness.
    \textbf{(c)} shows that the rate with which applications were granted remained constant before 2019. This explains why less research competitive universities were able to secure greater level of grant income post 2015.}
    \label{fig:fs_rob}
\end{figure*}

Fig. \ref{fig:fs_rob} is based on data for the years 2015-2019 released by UKRI. This data contains details on grant applications that did not receive funding and allows us to examine how universities responded to the relaxation of austerity after the election in 2015. Fig. \ref{fig:fs_rob}a and b show that the least research competitive universities increased the number and size of their grant applications most in response to the 2015 easing of austerity. Recent results on contest \cite{MS06,FNS20,OS20} suggest that making the distribution of prizes more symmetric will encourage efforts of weaker contestants. The data shows that the number of grants and the volume of grant applications submitted by less research competitive universities in 2017 increased more relative to 2015 than for the more research competitive universities. These additional results support our main findings. Grant income increases of less research competitive universities are the result of more applications being made by them and of greater levels of funding being requested after 2015.

\begin{figure*}[ht!]
  \centering
  \begin{subfigure}[t]{0.48\textwidth}
  \begin{subfigure}[t]{0.02\textwidth}
    \textbf{\textsf{a}}
  \end{subfigure}
  \begin{subfigure}[t]{0.9\textwidth}
    \includegraphics[width=\linewidth,valign=t]{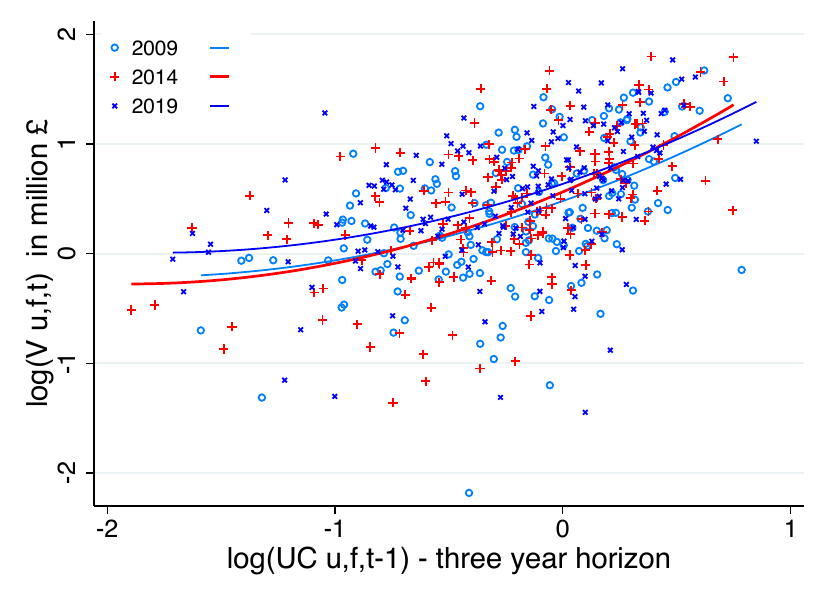}
  \end{subfigure}\hfill \\
  \begin{subfigure}[t]{0.02\textwidth}
    \textbf{\textsf{b}}
  \end{subfigure}
  \begin{subfigure}[t]{0.9\textwidth}
     \includegraphics[width=\linewidth,valign=t]{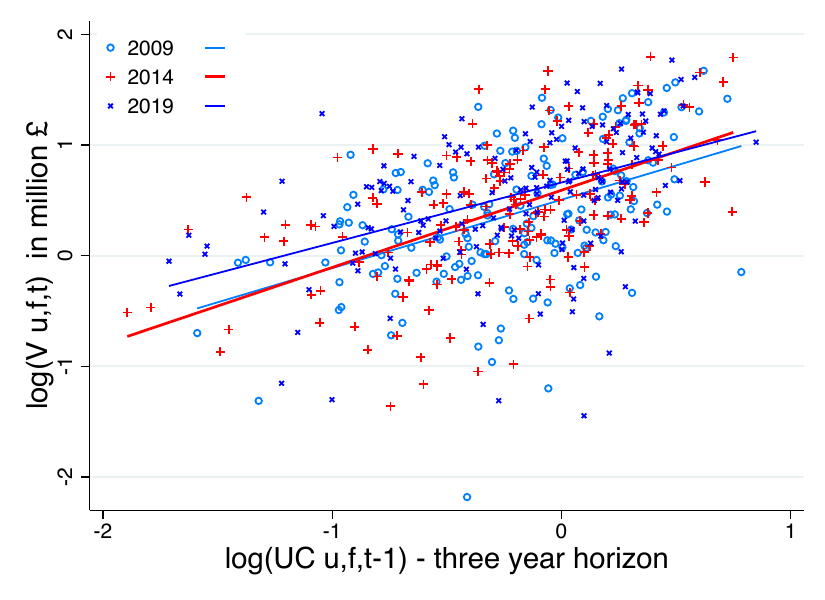}
  \end{subfigure}
  \end{subfigure}
  \begin{subfigure}[t]{0.48\textwidth}
  \begin{subfigure}[t]{0.02\textwidth}
    \textbf{\textsf{c}}
  \end{subfigure}
  \begin{subfigure}[t]{0.9\textwidth}
    \includegraphics[width=\linewidth,valign=t]{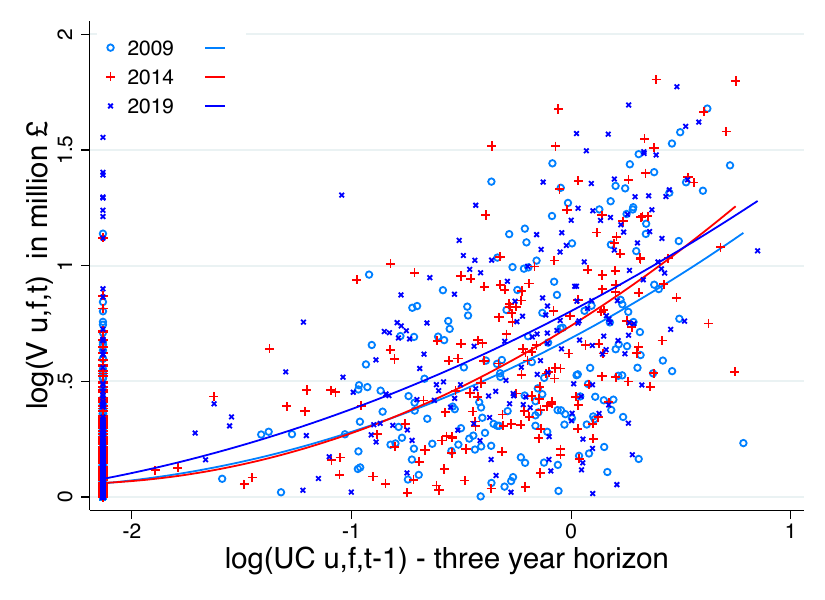}
  \end{subfigure}\hfill \\
  \begin{subfigure}[t]{0.02\textwidth}
    \textbf{\textsf{d}}
  \end{subfigure}
  \begin{subfigure}[t]{0.9\textwidth}
     \includegraphics[width=\linewidth,valign=t]{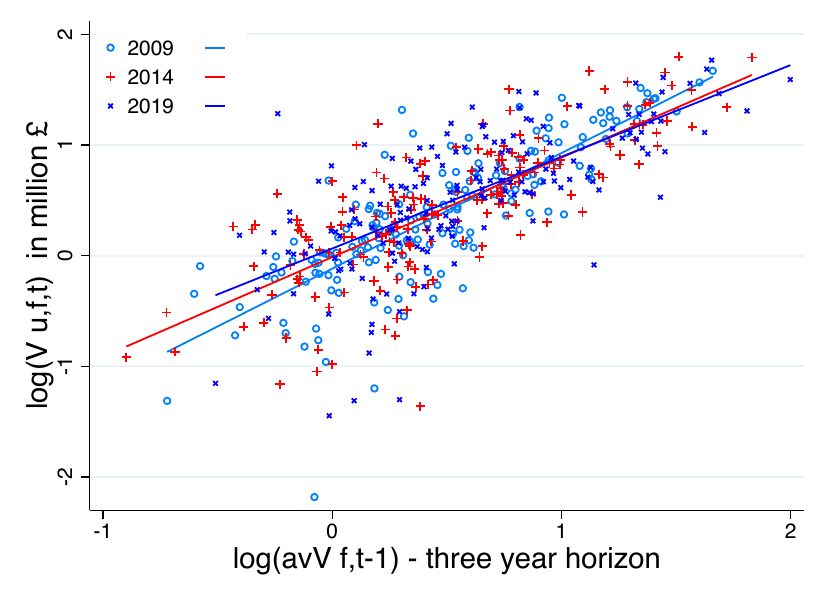}
  \end{subfigure}
  \end{subfigure}
    \captionsetup{justification=justified}
    \caption{\textbf{Robustness analysis for Figure \ref{fig:three} b}. \textbf{(a)} reproduces Figure \ref{fig:three} \textbf{(b)} provides the same analysis with a linear fit. \textbf{(c)} includes all observations with no grant income in a given year, by plotting $\log(V_{t,f}+1)$. All three versions indicate that the elasticity of university grant income w.r.t university competitiveness was greater under austerity that before or afterwards. \textbf{(d)} replaces university competitiveness with the average grant income of the previous three year period. The elasticity of grant income w.r.t. the average of lagged grant incomes appears to be getting smaller as time passes. }
    \label{fig:three_rob}
\end{figure*}

Fig. \ref{fig:three_rob} provides additional detail for panel b of Fig. \ref{fig:three}. We show that linear and quadratic fits to the data produce the same result: the slope of the correlation between university competitiveness (UC$_{u,f,t-1}$) and grant income (V$_{u,f,t}$) is steeper for the austerity years ending in 2014, than in the pre-austerity years ending in 2009 or the post austerity years ending in 2019. We also demonstrate that this finding is robust to inclusion of data points for which there is no grant income in a given year and for which the algorithm produces the minimal competitiveness score. Panel d of the figure shows that average grant income over the same three year periods has a quite different correlation with current grant income in this period: the slope of the relationship becomes increasingly flatter, suggesting weaker correlation between average lagged grant incomes and current grant income.

\begin{figure*}[ht!]
  \centering
  \begin{subfigure}[t]{0.02\textwidth}
    \textbf{\textsf{a}}
  \end{subfigure}
  \begin{subfigure}[t]{0.43\textwidth}
    \includegraphics[width=\linewidth,valign=t]{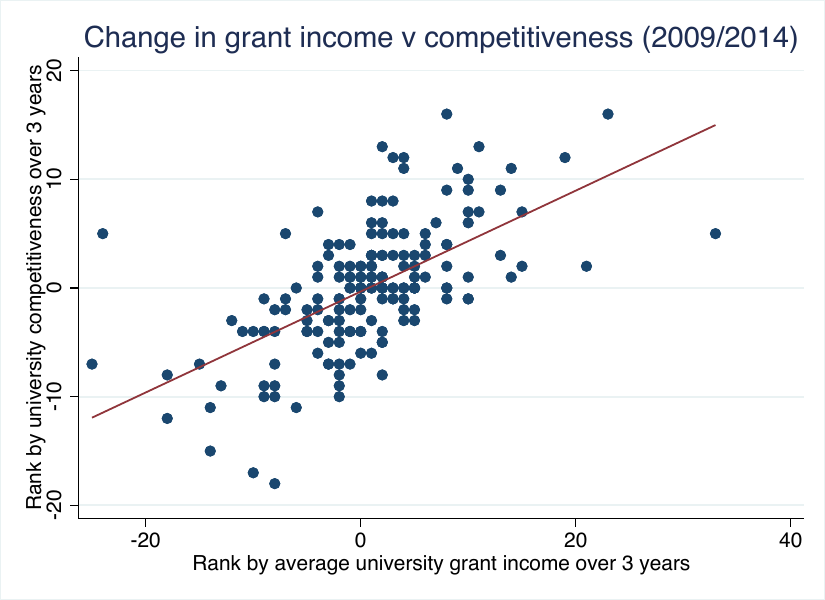}
  \end{subfigure}
  \begin{subfigure}[t]{0.02\textwidth}
    \textbf{\textsf{b}}
  \end{subfigure}
  \begin{subfigure}[t]{0.43\textwidth}
     \includegraphics[width=\linewidth,valign=t]{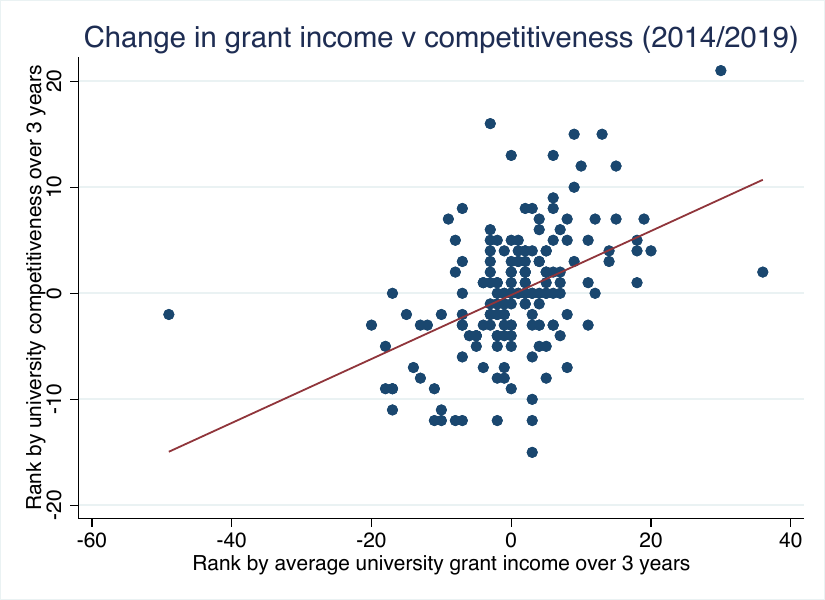}
  \end{subfigure}
  \captionsetup{justification=justified}
    \caption{\textbf{Rankings 2009-2019}. We compare changes in ranking position over 5 years (e.g. rank 2014 - rank 2009) of universities by average grant income and university competitiveness. Rankings are constructed using data for two years prior and the year listed, e.g. 2007-2009 for 2009. Both panels show that changes in ranking position based on grant income and changes in ranking position based on the competitiveness measure are positively correlated, but that the ranking changes generally differ. Generally a change in ranking by income translates into a smaller change in ranking according to the competitiveness measure. \textbf{(a)} Shows the change in ranking positions for 2014 relative to 2009. \textbf{(b)} Shows the change in ranking positions for 2019 relative to 2014. }
    \label{fig:rank}
\end{figure*}

Fig. \ref{fig:rank} shows that rankings by grant income and rankings by the university competitiveness measure are correlated but that there is significant heterogeneity across these measures, especially for the universities with highest and lowest ranks.

\section{Econometric Analysis of Austerity}
\label{sec:si_eres}

This section contains results on the effects of austerity for university grant income after the 2015 general election in the UK. The data available from UKRI on grant funding comprises the years 2006-2020 inclusive. 

\begin{table}[ht!]
{\small
\begin{longtable}{r | l l }
     \caption{ {\large \label{tab:vardes}
 Variable Descriptions} } \\
        & Variable & Description   \\ \hline 
        (1) & V$_{u,f,t}$                       &   University grant income by research council    \\
        (2)  & $\widebar{\text{V}}_{u,f,t}$           &    Average grant income by research council, three year window   \\
        (3)  & NG$_{u,f,t}$                      &   Number of grants by research council ($_f$)    \\
        (4) & UC$_{u,t}$                        & University competitiveness across research councils, three year window      \\
        (5) & UC$_{u,f,t}$                      &   University competitiveness by research council, three year window     \\
        (6) & $\widetilde{\text{UC}}_{u,t}$     &  University competitiveness across research councils, five year window      \\
        (7) & $\widetilde{\text{UC}}_{u,f,t}$   &   University competitiveness by research council, five year window    \\
        (8) & D$_{T}$                           &  Treatment dummy for NERC, AHRC, ESRC and EPSRC     \\
        (9) & D$_{Post}$                        &   Treatment period, post 2015     \\
        (10) & HHI$_{f,t}$                      &   Herfindahl-Hirschman Index of grant income concentration by research council    \\
        (11) & md V$_{f,t}$                     &  Median grant size by research council     \\
        (12) & $\Sigma \text{V}_{f,t}$                 &  Total grant income by research council     \\
        (13) & R$_{u,f,t}$                        &  Total grant count by research council     \\
\end{longtable}
}
\end{table}

\newpage 
In this period there were two general elections in May 2010 and May 2015. We split the UKRI data into three periods, corresponding to each of the governments in power between 2006 and 2020. The first government of David Cameron and George Osborne, in power from 2010 to 2015, introduced austerity and froze research budgets in nominal terms \cite{JSS19}. This meant that most research councils were faced with a real terms cut of research funding. Due to the lags inherent in the construction of the university competitiveness metrics we can only analyse the impact of the 2015 election, after which Cameron and Osborne allowed research budgets to increase with inflation, relaxing austerity. 

From the UKRI data we construct a number of variables, definitions of which are provided in Table \ref{tab:vardes}. 
 To analyse the effect of research competitiveness on universities' grant incomes we construct time series of  university competitiveness metrics using data over three (UC$_{u,f,t}$;UC$_{u,t}$) or five year periods ($\widetilde{\text{UC}}_{u,f,t}$; $\widetilde{\text{UC}}_{u,t}$), prior to the focal year $t$. The metrics are added to a balanced panel for 132 universities and research centers in the UK that received grants from UK research councils between 2006 and 2020.

 \begin{table*}[ht!]
\linespread{1} 
\centering
\caption{\large Basic information for seven research councils. }
\begin{tabular}{p{7cm} p{1.1cm}<{\centering} p{1.1cm}<{\raggedright} p{1.9cm}<{\centering}p{1.1cm}<{\centering}p{1.6cm}<{\centering}}
\toprule
\hline
Research Council & Abbr. & $N_{grant}$ & $Value (\times 10^{8})$ & $N_{subject}$ & $N_{institution}$\\ 
\hline
Arts and Humanities Research Council
& AHRC  &  $4,796$  & $8.8$    & $63$    & $168$ \\
Biotechnology and Biological Sciences Research Council
& BBSRC &  $4,661$  & $21.1$    & $54$    & $135$ \\
Engineering and Physical Sciences $\qquad$ Research Council 
& EPSRC  &  $13,888$  & $79.4$    & $62$    & $182$ \\
Economic and Social Research Council 
& ESRC &  $5,440$ & $24.3$   & $68$    & $205$ \\
Medical Research Council 
&  MRC  &  $5,216$  & $41.6$    & $22$    & $183$ \\
Natural Environment Research Council 
& NERC  &  $6,546$  & $18.6$   & $71$    & $201$ \\
Science and Technology Facilities Council
& STFC  &  $2,883$  & $16.2$    & $58$    & $95$ \\
\bottomrule
\end{tabular}
\label{tab:grant}
\end{table*}

 BBSRC is excluded from our analysis due to lack of comparable data for the years 2006-2010, leaving 132$\times$6=792 possible observations per year. Due to complete inactivity of some universities at some research councils there are in practice only 614 university-research council pairs. Metrics of university competitiveness are constructed using at least three years of data and given the need to lag the metrics by one year we have ten years of data with which to estimate the statistical models reported below. Table \ref{tab:ds} provides descriptive statistics and correlations of the variables in our data. Variables are discussed further in the Section on \hyperref[sec:m]{Methods}.

\begin{table*}[ht!]
\begin{ThreePartTable}
    \begin{TableNotes}
    {\scriptsize
    \item[1] Statistics: mn is mean, sd is standard deviation, md is median, m is minimum, M is maximum.
    \item[2] Indices: university (u), research council or funder (f), start year of grant (t).
    \item[3] V$_{u,f,t}$, $\widebar{\text{V}}_{f,t-1}$ and md V$_{f,t}$ are measured in million UK $\pounds$. HHI is multiplied by 1000,  $\Sigma \text{V}_{f,t}$ is divided by 100.
    \item[4] The columns denoted (1)-(12) contain Pearson's correlation coefficients.}
    \end{TableNotes}
\footnotesize{    
\begin{longtable}{r l*{2}{d{1.1}}
r@{\hspace{1\tabcolsep}}r@{\hspace{1\tabcolsep}}r@{\hspace{1\tabcolsep}} l*{12}{d{1.1}}}
    \caption{\normalsize \label{tab:ds} Descriptive Statistics } \\
      \toprule
  \multicolumn{2}{l}{N=6140}       &\multicolumn{18}{c}{} \\
       \multicolumn{2}{l}{}   &\multicolumn{1}{c}{mn} & \multicolumn{1}{c}{sd} &   \multicolumn{1}{c}{md} &     \multicolumn{1}{c}{m} &     \multicolumn{1}{c}{M} &      \multicolumn{1}{c}{(1)} &      \multicolumn{1}{c}{(2)} &      \multicolumn{1}{c}{\,\;(3)} &      \multicolumn{1}{c}{\,\;(4)} &      \multicolumn{1}{c}{\,\;(5)} &      \multicolumn{1}{c}{\,\;(6)} &      \multicolumn{1}{c}{\,\;(7)} &      \multicolumn{1}{c}{\,\;(8)} &      \multicolumn{1}{c}{\,\;\,(9)} &     \multicolumn{1}{c}{(10)} &     \multicolumn{1}{c}{(11)} &
          \multicolumn{1}{c}{(12)} \\
\hline
(1) & V$_{u,f,t}$ &     2.39&     7.17&        0&        0&      218&         &         &         &         &         &         &         &         &         &         &         &             \\
(2) & $\widebar{\text{V}}_{u,f,t-1}$ &     2.24&     6.35&        0&        0&       137&     0.72&         &         &         &         &         &         &         &         &         &         &              \\
(3) & NG$_{u,f,t}$ &     4.02&     7.17&        1&        0&       81&     0.79&     0.79&         &         &         &         &         &         &         &         &         &             \\
(4) & UC$_{u,t}$&     0.70&     1.39&        0&        0&       10&     0.49&     0.55&     0.65&         &         &         &         &         &         &         &         &           \\
(5) & UC$_{u,f,t}$&     0.29&     0.70&        0&        0&        7&     0.59&     0.69&     0.73&     0.74&         &         &         &         &         &         &         &          \\
(6) & $\widetilde{\text{UC}}_{u,t}$&     0.95&     2.01&        0&        0&       15&     0.49&     0.54&     0.63&     0.98&     0.72&         &         &         &         &         &         &              \\
(7) & $\widetilde{\text{UC}}_{u,f,t}$&     0.48&     1.22&        0&        0&       22&     0.44&     0.52&     0.56&     0.62&     0.72&     0.61&         &         &         &         &         &               \\
(8) & D$_{T}$ &     0.72&     \multicolumn{1}{c}{$-$} &        1&        0&        1&     0.00&    -0.01&     0.05&    -0.06&     0.02&    -0.06&     0.07&         &         &         &         &             \\
(9) & D$_{Post}$   &     0.50&     \multicolumn{1}{c}{$-$} &        0&        0&        1&     0.03&     0.04&    -0.01&    -0.00&    -0.00&     0.01&     0.01&     0.00&         &         &         &                \\
(10) & HHI$_{f,t}$  &    55.77&    19.97&       52&       30&      117&     0.01&     0.02&    -0.02&     0.00&    -0.01&     0.01&    -0.01&    -0.20&     0.10&         &         &        \\
(11) & md V$_{f,t}$&     0.24&     0.15&        0&        0&        1&     0.15&     0.16&     0.08&    -0.01&     0.00&    -0.01&    -0.05&    -0.42&     0.19&     0.20&         &                 \\
(12) & $\Sigma \text{G}_{f,t}$&     4.40&     2.50&        3&        2&       11&     0.19&     0.21&     0.20&    -0.05&     0.04&    -0.05&    -0.01&     0.36&    -0.09&    -0.16&     0.25&                 \\
(13) & R$_{u,t}$ &    21.66&    34.25&        4&        0&      209&     0.42&     0.46&     0.60&     0.89&     0.67&     0.86&     0.59&    -0.07&    -0.02&     0.01&    -0.05&    -0.09 \\  
\bottomrule
\insertTableNotes    
    \end{longtable}}
\end{ThreePartTable}
\end{table*}

Table \ref{tab:ols} sets out seven models pooling all the data, in which we control for start-year, university and research council fixed effects as well as the concentration of lagged grant funding, the number of grants and the median value of grants approved by the research council in the previous year. We report coefficients for UC$_{u,t-1}$ and UC$_{u,f,t-1}$. 

{\small
\begin{table*}[htp!]
\begin{ThreePartTable}
    \begin{TableNotes}
    {\footnotesize
    \item[1] Robust standard errors are reported in parentheses:
      $^{*}$ $(p < 0.05)$, $^{**}$ $(p < 0.01)$,
      $^{***}$ $(p < 0.001)$.
    \item[2] All models contain year fixed effects.
    \item[3] Models 1 and 2 report results aggregated across councils by year.
    \item[4] Models 3-7 contain funder fixed effects and controls for concentration of grant funding in the previous year, number of grants awarded in the previous year and median grant value.
    \item[5] Column 6 presents results for years prior to 2016. Column 7 presents results for years after 2015.
    \item[7] Coefficients for competitiveness measures not significant with university-funder fixed effects. Model not reported here.}
    \end{TableNotes}
\begin{longtable}[htbp]{l | l*{1}{d{3.5}} | l*{2}{d{3.5}} | l*{2}{d{3.5}}}
    \caption{\normalsize \label{tab:ols} Pooled Poisson Models for Annual University Grant Income (V$_{u,f,t}$) } \\
      \toprule   
      &\multicolumn{1}{c}{(1)} & \multicolumn{1}{c|}{(2)} & \multicolumn{1}{c}{(3)} & \multicolumn{1}{c}{(4)} & \multicolumn{1}{c|}{(5)}&\multicolumn{1}{c}{(6)} & \multicolumn{1}{c}{(7)} \\
\toprule   
UC$_{u , t-1}$ & 0.418$^{***}$ & -0.023 & -0.018 & & -0.156^{***} & -0.186$^{***}$ & -0.258^{**} \\
               & (0.037) & (0.030) & (0.027) & & (0.038)  &  (0.055)  & (0.096)  \\
UC$_{u ,f ,t-1}$ &                     &                     &                     &       0.281^{***} &       0.325^{***} &       0.358$^{***}$ &       0.301^{***} \\
                    &                     &                     &                     &     (0.046)         &     (0.045)         &     (0.049)         &     (0.064)         \\
University f.e.     & \multicolumn{1}{c}{No}  & \multicolumn{1}{c|}{Yes} & \multicolumn{1}{c}{Yes} & \multicolumn{1}{c}{Yes} & \multicolumn{1}{c|}{Yes} & \multicolumn{1}{c}{Yes} & \multicolumn{1}{c}{Yes}  \\
\hline
Observations        &    \multicolumn{2}{c|}{1280}         &        \multicolumn{3}{c|}{6140}           &        \multicolumn{2}{c}{3070}  \\
Log-likelihood                  &     \multicolumn{1}{c}{-10,351}         &      \multicolumn{1}{c|}{-3,211}         &    
\multicolumn{1}{c}{ -11,013}         &     \multicolumn{1}{c}{-10,486}         &     \multicolumn{1}{c|}{-10,405}         &      \multicolumn{1}{c}{-4,286}         &      \multicolumn{1}{c}{-5,705}         \\
\bottomrule 
      \insertTableNotes    
    \end{longtable}
  \end{ThreePartTable}
\end{table*}
}

\small{
\begin{table*}    
\begin{ThreePartTable}
    \begin{TableNotes}
  {\footnotesize
    \item[1] Robust standard errors, clustered on university and research council, in parentheses:
      $^{*}$ \(p<0.05\), $^{**}$ \(p<0.01\),
      $^{***}$ \(p<0.001\).
    \item[2] All models include research council(f), university(u) and time(t) fixed effects. Models 2,4-6 include HHI$_{f,t}$, md V$_{u,f,t-1}$, V$^2_{u,f,t-1}$, NG$_{f,t-1}$, NG$_{f,t-2}$ and R$_{f,t}$.
    \item[3] Model 5 contains research council time trends interacted with D$_{Post}$.}
    \end{TableNotes}
    \begin{longtable}[htbp]{ l*{6}{d{5.4}}}
    \caption{\hspace*{-0.15cm} \normalsize \label{tab:g} Main Results - Fixed Effects Models for University Grant Income (V$_{u,f,t}$)} \\
      \toprule
 $N=6140$   &\multicolumn{1}{c}{(1)}&\multicolumn{1}{c}{(2)}&\multicolumn{1}{c}{(3)}&\multicolumn{1}{c}{(4)}&\multicolumn{1}{c}{(5)}&\multicolumn{1}{c}{(6)} \\
\toprule 
D$_{T} \times \text{D}_{Post}$ &     0.393^{***}     &   0.257^{*}             &                     &                     &                     &                     \\
                    &     (0.081)         &      (0.106)         &                     &                     &                     &                     \\
D$_{Post} \times \text{UC}_{u,f,t-1}$ &                     &                     &      -0.111^{***}&      -0.106^{***}&      -0.111^{**} &                     \\
                    &                     &                     &     (0.029)         &     (0.029)         &     (0.033)         &                     \\
2012 $\times \text{UC}_{u,f,t-1}$ &                     &                     &                     &                     &                     &      -0.010         \\
                    &                     &                     &                     &                     &                     &     (0.032)         \\
2013 $\times \text{UC}_{u,f,t-1}$&                     &                     &                     &                     &                     &      -0.018         \\
                    &                     &                     &                     &                     &                     &     (0.060)         \\
2014 $\times \text{UC}_{u,f,t-1}$&                     &                     &                     &                     &                     &      -0.067         \\
                    &                     &                     &                     &                     &                     &     (0.049)         \\
2015 $\times \text{UC}_{u,f,t-1}$&                     &                     &                     &                     &                     &      -0.076         \\
                    &                     &                     &                     &                     &                     &     (0.073)         \\
2016 $\times \text{UC}_{u,f,t-1}$&                     &                     &                     &                     &                     &      -0.076  \\
                    &                     &                     &                     &                     &                     &     (0.061)         \\
2017 $\times \text{UC}_{u,f,t-1}$&                     &                     &                     &                     &                     &      -0.145^{***} \\
                    &                     &                     &                     &                     &                     &     (0.044)         \\
2018 $\times \text{UC}_{u,f,t-1}$&                     &                     &                     &                     &                     &      -0.199^{*} \\
                    &                     &                     &                     &                     &                     &     (0.081)         \\
2019 $\times \text{UC}_{u,f,t-1}$&                     &                     &                     &                     &                     &      -0.194^{*} \\
                    &                     &                     &                     &                     &                     &     (0.087)         \\
2020 $\times \text{UC}_{u,f,t-1}$&                     &                     &                     &                     &                     &      -0.144^{*} \\
                    &                     &                     &                     &                     &                     &     (0.060)         \\
Time trends & \multicolumn{1}{c}{No} & \multicolumn{1}{c}{No} & \multicolumn{1}{c}{No} & \multicolumn{1}{c}{No} & \multicolumn{1}{c}{Yes} & \multicolumn{1}{c}{No} \\       
\hline
Log-likelihood                 &      \multicolumn{1}{c}{-6,420}         &      \multicolumn{1}{c}{-6,350}         &      \multicolumn{1}{c}{-6,331}         &      \multicolumn{1}{c}{-6,322}         &      \multicolumn{1}{c}{-6,229}         &      \multicolumn{1}{c}{-6,286}         \\
      \midrule
 \insertTableNotes   
    \end{longtable}
  \end{ThreePartTable}
\end{table*}
}

{\normalsize
The specifications reported in Table \ref{tab:ols} still leave open the possibility that unobserved university-research council fixed effects explain the positive correlation between UC$_{u,t}$ and grant income. This motivates estimating difference-in-differences models, exploiting the 2015 relaxation of austerity. We define D$_{T}$ as all applications made at NERC, AHRC, ESRC and EPSRC; second we define UC$_{f,2015}$ as university competitiveness at research council level for 2013-2015; and third we use a sliding university competitiveness (UC$_{u,f,t}$) index interacted with year dummies. 

Table \ref{tab:g} provides estimates of $\delta$ in Equation \ref{eq:model}. As a point of comparison the models in Columns (1) and (2) provide treatment effects at the research council level. These models are based on the observation that grant funding directed to NERC, AHRC, ESRC and EPSRC increased particularly strongly in 2015. The estimated coefficients confirm that this increased grant incomes for applicants relative to applications directed to the remaining research councils. However, Column (2) illustrates that these effects are less likely to be significant and smaller, once we include university and research council level controls in the model. Columns 3-6 show that the university competitiveness measure UC$_{u,f,t-1}$ reflecting the ranking of universities by research council identifies the predicted negative treatment effect of competitiveness on university grant income after the 2015 election. }

\begin{table*}[ht!]
\begin{ThreePartTable}
    \begin{TableNotes}
    {\footnotesize
    \item[1] All models correspond to those reported in Table \ref{tab:g}.}
    \end{TableNotes}
    \begin{longtable}{ l*{4}{d{7.7}}}
    \caption{\large \label{tab:ga} Fixed Effects Models for Annual University Grant Income (V$_{u,f,t}$) \\ \centering Alternative Metric } \\
      \toprule
 $N=6140$   &\multicolumn{1}{c}{(3)}&\multicolumn{1}{c}{(4)}&\multicolumn{1}{c}{(5)}&\multicolumn{1}{c}{(6)}\\
\toprule 
D$_{Post} \times \Bar{\text{V}}_{u,f,t-1} $                    &      0.000 &      0.001 &      -0.003&                     \\
                                        &     (0.007)         &     (0.007)         &     (0.007)         &                     \\
2012 $\times \Bar{\text{V}}_{u,f,t-1}$                   &                     &                     &                     &      -0.008         \\
                               &                     &                     &                     &     (0.005)         \\
2013 $\times \Bar{\text{V}}_{u,f,t-1}$                    &                     &                     &                     &       -0.014^{*}          \\
                                   &                     &                     &                     &     (0.007)         \\
2014 $\times \Bar{\text{V}}_{u,f,t-1}$                    &                     &                     &                     &      -0.017^{**}  \\
                                   &                     &                     &                     &     (0.005)         \\
2015 $\times \Bar{\text{V}}_{u,f,t-1}$                  &                     &                     &                     &       -0.016   \\
                                       &                     &                     &                     &     (0.008)         \\
2016 $\times \Bar{\text{V}}_{u,f,t-1}$       &                     &                     &                     &      -0.011 \\
                   &                     &                     &                     &     (0.010)         \\
2017 $\times \Bar{\text{V}}_{u,f,t-1}$                  &                     &                     &                     &       -0.012   \\
                            &                     &                     &                     &     (0.009)         \\
2018 $\times \Bar{\text{V}}_{u,f,t-1}$                &                     &                     &                     &      -0.013\\
                       &                     &                     &                     &     (0.012)         \\
2019 $\times \Bar{\text{V}}_{u,f,t-1}$             &                     &                     &                     &      -0.013\\
                        &                     &                     &                     &     (0.012)         \\
2020 $\times \Bar{\text{V}}_{u,f,t-1}$                   &                     &                     &                     &      -0.016 \\
                &                     &                     &                     &     (0.010)         \\
\hline
Log-likelihood                 &      
      \multicolumn{1}{c}{-6,267}         &      \multicolumn{1}{c}{-6,256 }         &      \multicolumn{1}{c}{-6,181}         &      
\multicolumn{1}{c}{-6,236} \\
      \midrule
      \insertTableNotes    
    \end{longtable}
  \end{ThreePartTable}
\end{table*}  

{\normalsize
In column 4 we introduce the HHI over lagged grant income at research council level as a further control variable to demonstrate that UC$_{u,f,t}$ is not a proxy for concentration of grant income \cite{cH21}. In Column 5 we show that the treatment effect of the 2015 election is robust to research council time trends interacted with the policy dummy. In Column 6 we interact UC$_{u,f,t}$ with time dummies and find that this interaction becomes significant after 2015. The specification rules out that events preceding this year or following in later years are the cause of the policy shock of 2015 \cite{dhA03}. We report the coefficients from Column 6 in Fig.~\ref{fig:three}c. All models provided in Table \ref{tab:g} indicate that conditional on university-research council fixed effects, those universities ranked higher on the complexity measure lost grant income shares to institutions ranked lower down after 2015. 
}

{\small
\begin{table*}[ht!]
\begin{ThreePartTable}
    \begin{TableNotes}
    {\footnotesize
    \item[1] Robust standard errors, clustered on university and research council, in parentheses:
      $^{*}$ \(p<0.05\), $^{**}$ \(p<0.01\),
      $^{***}$ \(p<0.001\).
    \item[2] The table reports coefficients of Poisson DD estimates of the effects of austerity ending on the annual grant income of UK universities at research council level. D$_{Post}=1$ after 2015.  
    \item[3] All models include research council(f), university(u) and time(t) fixed effects. Furthermore all models include  a quadratic function of lagged median grant income at research council level, interacted one and two period lags of total grants awarded at research council level and the lagged count of grants awarded to the focal university at other research councils.
    \item[4] Model 1 replicates column 4 from Table \ref{tab:g}. In Column 2 we restrict the sample to the years 2014-2017. In Column 3 we replace the University Competitiveness measure based on a three year window with a measure based on a five year window. Both of these models are estimated using Poisson fixed effects models. In Columns 4-6 we present results from estimating linear fixed effects models; here the dependent variable is the logarithm of university grant income. The number of observations is reduced as all cases in which grant income is zero are removed from the sample.
    \item[5] Models 5 and 6 contains research council time trends interacted with D$_{Post}$. }
    \end{TableNotes}
    \begin{longtable}{ l*{3}{d{2.4}} | l*{4}{d{2.4}}}
    \caption{\normalsize \label{tab:gir} Fixed Effects Models for Annual University Grant Income (V$_{u,f,t}$) - Robustness} \\
      \toprule
                    &\multicolumn{3}{c|}{\text{Alternative specifications}}&\multicolumn{4}{c}{\text{Linear fixed effects models}}\\
                    &\multicolumn{1}{c}{(1)}&\multicolumn{1}{c}{(2)}&\multicolumn{1}{c|}{(3)}& &\multicolumn{1}{c}{(4)}&\multicolumn{1}{c}{(5)}&\multicolumn{1}{c}{(6)}\\
\hline
D$_{Post} \times \widetilde{\text{UC}}_{u,f,t-1}$ &      -0.106^{***}&      -0.077^{*}  &                     & D$_{Post} \times \text{UC}_{u,f,t-1}$ &      -0.102^{***}&      -0.099^{**} &                     \\
                    &     (0.029)         &     (0.039)         &                     & &    (0.030)         &     (0.030)         &                     \\
2012 $\times \widetilde{\text{UC}}_{u,f,t-1}$ &                     &                     &      -0.036         & 2012 $\times \text{UC}_{u,f,t-1}$ &                    &                     &       0.040         \\  
                    &                     &                     &     (0.032)         & &                    &                     &      (0.068)         \\                 
2013 $\times \widetilde{\text{UC}}_{u,f,t-1}$ &                     &                     &      -0.037         & 2013 $\times \text{UC}_{u,f,t-1}$ &                     &                     &      -0.009         \\  
                    &                     &                     &     (0.062)         & &                    &                     &      (0.073)         \\                 
2014 $\times  \widetilde{\text{UC}}_{u,f,t-1}$ &                     &                     &      -0.093         & 2014 $\times \text{UC}_{u,f,t-1}$ &                   &                     &       0.006         \\  
                    &                     &                     &     (0.057)         & &                    &                     &      (0.062)         \\                 
2015 $\times  \widetilde{\text{UC}}_{u,f,t-1}$ &                     &                     &      -0.085         & 2015 $\times \text{UC}_{u,f,t-1}$ &                   &                     &      -0.025         \\  
                    &                     &                     &     (0.078)         &  &                   &                     &      (0.065)         \\                 
2016 $\times  \widetilde{\text{UC}}_{u,f,t-1}$ &                     &                     &      -0.088         & 2016 $\times \text{UC}_{u,f,t-1}$ &                   &                     &      -0.011         \\  
                    &                     &                     &     (0.057)         &  &                   &                     &      (0.068)         \\                 
2017 $\times  \widetilde{\text{UC}}_{u,f,t-1}$ &                     &                     &      -0.175^{***} & 2017 $\times \text{UC}_{u,f,t-1}$ &                     &                     &      -0.116^{*}  \\  
                    &                     &                     &     (0.043)         &  &                   &                     &      (0.059)         \\                 
2018 $\times  \widetilde{\text{UC}}_{u,f,t-1}$ &                     &                     &      -0.195^{*}  &  2018 $\times \text{UC}_{u,f,t-1}$  &                  &                     &      -0.212^{**} \\  
                    &                     &                     &     (0.086)         &   &                  &                     &      (0.075)         \\                 
2019 $\times  \widetilde{\text{UC}}_{u,f,t-1}$ &                     &                     &      -0.217^{*}  &  2019 $\times \text{UC}_{u,f,t-1}$ &                   &                     &      -0.048         \\  
                    &                     &                     &     (0.086)         &  &                   &                     &      (0.069)         \\                 
2020 $\times  \widetilde{\text{UC}}_{u,f,t-1}$ &                     &                     &      -0.165^{*}  & 2020 $\times \text{UC}_{u,f,t-1}$ &                   &                     &      -0.114         \\  
                    &                     &                     &     (0.064)         &  &                   &                     &      (0.062)         \\ 
Time trends & \multicolumn{1}{c}{No} & \multicolumn{1}{c}{No} & \multicolumn{1}{c|}{No} & & \multicolumn{1}{c}{No} & \multicolumn{1}{c}{Yes} & \multicolumn{1}{c}{Yes} \\       
\hline
  Observations        &        \multicolumn{1}{c}{6140}         &        \multicolumn{1}{c}{2072}         &        \multicolumn{1}{c|}{5445}         &  &      \multicolumn{3}{c}{3884}       \\
  Log-likelihood            &             \multicolumn{1}{c}{-6,305}  &       \multicolumn{1}{c}{ -1,903}  &         \multicolumn{1}{c|}{-6,280}  &   &     \multicolumn{1}{c}{-5,289}   &       \multicolumn{1}{c}{-5,277}    &      \multicolumn{1}{c}{-5,277}  \\  
      \midrule
      \insertTableNotes    
    \end{longtable}
  \end{ThreePartTable}
\end{table*}  
}

{\normalsize
The findings suggest that the descriptive results provided in Fig.~\ref{fig:three}b are robust: the end of austerity reduced the relative advantage higher competitiveness provided to university grant applicants. 

This still leaves the question whether the particular metric we propose is required to measure university competitiveness using grant income? Hitherto most university rankings are based on an aggregate of grant income. Sometimes rankings are restricted to specific research fields in recognition of the fact that summing up grant income across disparate research areas is problematic \cite{GWB19}. 

To compare the performance of UC metrics and measures of aggregate grant income we estimate the models presented in Table \ref{tab:g} using the average of lagged grant income from the previous three years ($\Bar{V}_{u,f,t}$) as a measure of competitiveness. Table \ref{tab:ga} provides the results. The estimates of $\delta$ in this model using average lagged grant income are never significant for the models in Columns (3)-(5) of the table. The more flexible model set out in Column (6) illustrates that lagged aggregate grant income at university level is significant prior to 2015 and effects are constant over time. This illustrates that UC is a better measure of competitiveness, due to the dynamic weighting of research fields this measure contains.   

Finally, we set out several additional tests of the robustness of our principal findings. Table \ref{tab:gir} shows that reducing the number of years in the sample and increasing the number of years used for construction of the university competitiveness index does not change our results. It also demonstrates that estimating the models we present using OLS, which means discarding observations for years in which universities have no grant income, does not change any of the principal implications we derived here.
}  

\section{Funding Science and Evaluating Science Funding through Peer Review}
\label{app:fsci}

\normalsize{
This section sets out a framework for university research funding and the role of metrics within this. The main purpose of the section is to highlight the importance of the assumption that grants are awarded on the basis of quality for any statistical analysis relying on metrics derived from grant data.

Research grant funding by the state has a long history\cite{mB18,AL20}. The Royal Society (1849) and the Medical Research Council (1913) in the UK, the Notgemeinschaft in Germany (1920), the Caisse nationale des sciences in France (1930) and the National Institutes of Health in the United States (1946) are examples of institutions set up to channel government funds into research grants. Systematic measurement of scientific outputs begins with J.M. Catell, editor of Science in 1906 \cite{bG06}. Measurement became linked to evaluation of science funding in the late 1960s when citation data became available with the Science Citation Index (SCI). At this time the National Science Foundation (NSF) was first required to regularly report on the status of science \cite{aC20} and turned towards using citation data. In the UK financial pressures led to a first systematic assessment of grant funding in 1985 \cite{ajP89}. One outcome of the creation of the SCI was the derivation of the Journal Impact Factor (JIF) \cite{eG06} from that data. This measure was to become increasingly influential, leading to widespread use of rankings of scientists and universities based almost entirely on the JIF. 

Recently there have been a series of initiatives\cite{dora, LM15, dec} pushing back against the use of the JIF and related metrics to evaluate researchers and publications. These highly influential initiatives highlight a controversy around the use of metrics in the evaluation of science: it is argued that metrics will become distorted, if they have resource implications (Goodhart's Law \cite{mS96}) and that metrics cannot fully reflect the value of individual scientific contributions or individual scientists. Instead, it is argued, such evaluations should be based on peer review.  Calls for reform of the use of metrics in science evaluation are primarily informed by clear examples of dis-function \cite{yG16,bK20}. This literature lacks a framework setting out links between the science metrics and science funding. Drawing on the cycle of credibility \cite{yG20} that focuses on the role of peer review, we develop a graph of financial flows funding science, knowledge flows resulting from this and science metrics.

\begin{figure*}[htbp!]
     \centering
     \includegraphics[]{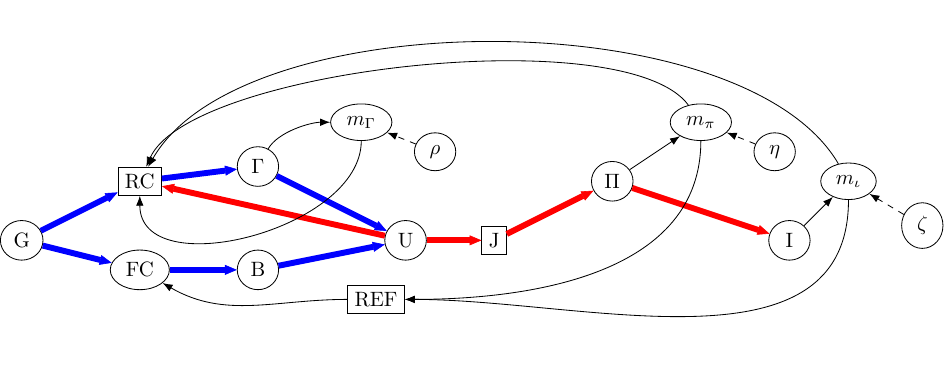}
     \captionsetup{justification=justified}
     \caption{Dual support: Financial flows (blue), knowledge flows (red) and measurement of outputs in the UK. Square nodes denote points at which peer review evaluation takes place. G: government; RC: research council; FC: Higher education funding council; B: block grant; $\Gamma$: grant income; U: university; J: journals;$\Pi$: publications; I: impact; REF: Research Excellence Framework.  $m_{\Gamma}$,$m_{\Pi},m_{\iota}$ are measures of grant income, publications and impact that characterise university $U$. $\eta,\rho,\zeta $ denote unobserved statistical error affecting the measures. Notice that, $m_{\Gamma}$ is forward-looking while $m_{\Pi},m_{\iota}$ are backward- looking\cite{iR18}.}
     \label{fig:ds}
 \end{figure*}

In the UK basic science research at universities is funded through a dual support mechanism\cite{N15}: a block grant is provided to each university by regional higher education funding councils, e.g. Research England, and researchers submit funding bids for specific grants to seven research councils: AHRC, ESRC, EPSRC, BBSRC, NERC, STFC, MRC. While grant applications are peer reviewed before funding is allocated the block grant is awarded for a period of seven years based on regular evaluations of the universities through a mechanism called the Research Excellence Framework (REF)\cite{gS17}. Fig. \ref{fig:ds} sets out financial flows, knowledge flows and three principal science metrics under the dual support system. The figure highlights the role of bibliometric measures ($m_{\Pi}$), grant income measures ($m_{\Gamma}$) and recently measures of impact ($m_{\iota}$) in informing evaluations of scientific output by research councils and higher-education funding councils. 

Three points emerge from Fig. \ref{fig:ds}: First, the central role of peer review within the system of science\cite{yG20}: research grants are evaluated at research councils, submitted papers are evaluated by journal editors and their reviewers and finally in the UK the REF is partly based on peer review. Metrics of grant income and publications draw on results of peer review, whereas metrics of impact are broader and may or may not draw on evaluation. Second, any grant income metric is forward-looking whereas the other indicators are backward-looking. This means grant income metrics capture current scientific competitiveness better than publication or impact metrics that generally become increasingly informative with the passage of time. Third, as long as there is selection for quality within the scientific system, the quality of grant applications, resulting publications and impacts will be correlated and so will the metrics that reflect each of these outputs. The literature finds evidence of such correlations \cite{GWB19}. 

Peer review of grants generally relies on status indicators as well as a review of the research proposal by experts, the peers. The Matthew effect \cite{rkM68, ASW14} may create distortions in grant allocation, if status indicators matter too much or if failure to win grant income discourages applicants from further bids \cite{BVR18}. In both instances competition for grant funding is weakened. Where status effects in grant allocation are very powerful, entry by new researchers is undermined and eventually science becomes less productive. Where status effects are not too strong the allocation of research grants should be on merit of each funding proposal. Then new methods, new research subjects and well trained new researchers can come together to create opportunities for funding to be allocated to applicants that are not well established and who do not already hold positions at high status universities. The evidence on dynamics of grant funding in the UK that we present suggests this is how UK grant funding currently operates.

The importance of such competition for our econometric analysis emerges from a simple model of grant income that draws on the characterisation of high quality research in Nurse's review of UK research councils \cite{N15}. Review ($R_i$) of a grant proposal $i$ by external reviewers determines whether a grant ($G_i$) is awarded. The outcome of this review is a function of the quality ($Q_{i}$) of a grant proposal and random unobserved factors $(\nu_i)$, such as the fit between the reviewers and the proposal, their ability to devote time to the review and other similar factors. The quality of a grant proposal is a function of three principal factors: the quality of the problem ($P_{i,j}$) the proposal focuses on, this is called "what" by Nurse\cite{N15} and usually also includes "how" the problem is to be addressed; the quality of the environment ($U_j$) within which the proposed research will be undertaken, called "where" by Nurse; the expertise of the applicants ($R_j$), called "who" by Nurse. Here the $j$ subscripts indicate university level effects. In addition, the quality of the proposal will depend on unobservable effort ($e_{i,j}$) and unobservable random factors ($\epsilon_i$). Fig. \ref{fig:one} presents this model as a directed acyclic graph. For simplicty the figure does not allow for co-determination of $P_{i,j}, U_j,R_j, e_{i,j}$ (PURe), but these factors are likely to be correlated in many cases.

\begin{figure}[hbtp!]
     \centering
     \includegraphics{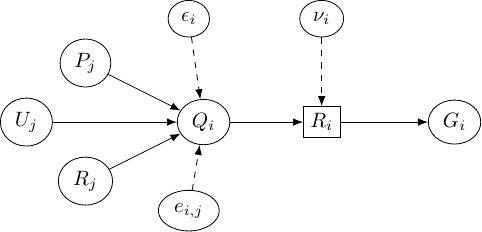}
     \caption{A model of research grant quality and evaluation}
     \label{fig:one}
\end{figure}

A metric that averages over all grants acquired by a university will capture the joint effect of the PURe factors characterising the university research environment. As long as incentives that elicit high levels of effort are in place, the PURe factors will be persistent over time in each university. Selection for high quality between universities will also tend to concentrate the strongest researchers in the same universities. 

Peer review of a grant usually focuses primarily on the quality of the research proposal, not on the status of applicants or of their institutions. In the context of significant uncertainty about benefits of each research grant peer review generates new information and enables competition, which elicits effort. In contrast, peer review becomes dis-functional, if assessment of the merits of a grant proposal is replaced explicitly or implicitly by metrics that reflect the status of a researcher or their institution. Then the Matthew effect comes into play \cite{rkM68,ASW14} and status trumps current effort. This points to a significant limitation of any science metric: it cannot in the long-run replace the process of evaluation. Any metric that contains information does so because robust evaluation and competition are in place. In the absence of robust competition for grant income econometric analysis of Equations such as \ref{eq:model} might still yield significant results, but these would be spurious. Therefore the evidence we provide on dynamics of grant income across research subjects and universities is an important complement to the regression results we report.

}

\bigskip

{\small
\begin{table*}
\linespread{1} 
\centering
\begin{tabular}{p{6.5cm}p{3cm}<{\centering} p{1.5cm}<{\centering}p{3cm}<{\centering}}
\toprule
University & Total Amount ($\pounds$) & $N_{subject}$ & Group\\ 
\hline
University of Edinburgh               & 755392270  & 101        & Russell Group \\
University of Leeds                   & 501470072  & 101        & Russell Group \\
University College London             & 1146544034 & 101        & Russell Group \\
University of Oxford                  & 1117494232 & 101        & Russell Group \\
University of Sheffield               & 430842015  & 100        & Russell Group \\
University of Cambridge               & 1012078241 & 99         & Russell Group \\
University of Birmingham              & 386800294  & 98         & Russell Group \\
University of Bristol                 & 504930773  & 98         & Russell Group \\
University of Nottingham              & 474058418  & 98         & Russell Group \\
Cardiff University                    & 317320465  & 96         & Russell Group \\
University of Southampton             & 507527952  & 96         & Russell Group \\
University of Glasgow                 & 469789863  & 95         & Russell Group \\
King's College London                 & 395823331  & 92         & Russell Group \\
University of Liverpool               & 372126937  & 92         & Russell Group \\
Imperial College London               & 1042091498 & 90         & Russell Group \\
Newcastle University                  & 366635812  & 90         & Russell Group \\
Queen's University of Belfast         & 189389905  & 86         & Russell Group \\
University of Manchester              & 835801582  & 102        & 1994 Group    \\
Queen Mary, University of London      & 217020953  & 94         & 1994 Group    \\
University of Warwick                 & 366110109  & 91         & 1994 Group    \\
University of York                    & 233740372  & 89         & 1994 Group    \\
University of Exeter                  & 231547824  & 88         & 1994 Group    \\
Durham University                     & 262598757  & 87         & 1994 Group    \\
Lancaster University                  & 162899012  & 85         & 1994 Group    \\
University of Leicester               & 151200485  & 83         & 1994 Group    \\
University of Sussex                  & 121709735  & 83         & 1994 Group    \\
University of Bath                    & 160569804  & 82         & 1994 Group    \\
University of Reading                 & 142664882  & 81         & 1994 Group    \\
University of Surrey                  & 151400425  & 78         & 1994 Group    \\
University of East Anglia             & 122769894  & 74         & 1994 Group    \\
Royal Holloway, University of London  & 74050579   & 74         & 1994 Group    \\
Loughborough University               & 168070726  & 70         & 1994 Group    \\
University of St Andrews              & 177300153  & 70         & 1994 Group    \\
University of Essex                   & 252708508  & 58         & 1994 Group    \\
Birkbeck College                      & 36282362   & 49         & 1994 Group    \\
London School of Economics \& Pol Sci & 86335816   & 43         & 1994 Group    \\
Goldsmiths College                    & 21271553   & 32         & 1994 Group    \\
School of Oriental \& African Studies & 21264774   & 29         & 1994 Group   \\
\bottomrule
\end{tabular}
\caption{\textbf{Research universities in Russell Group and $1994$ Group.} Total amount indicates the total amount of funding allocated to each university. $N_{subject}$ denotes the number of different research subjects in which universities obtain the grants. Note that here the universities in both groups have been classified as 1994 Group.}
\label{tab:universities}
\end{table*}
}

\end{document}